\documentclass[preprint]{elsarticle}
\usepackage[utf8]{inputenc}
\usepackage{amsmath}
\usepackage{amsfonts}
\usepackage{amssymb}
\usepackage{graphicx}
\usepackage{tabularx}
\usepackage{booktabs}
\usepackage{siunitx}
\usepackage{wrapfig}
\usepackage[version=3]{mhchem}
\usepackage{subfigure}
\usepackage{caption}	
\usepackage{fancyhdr}
\newenvironment{keywords}%
   {\begin{trivlist}\item[]{\bfseries\sffamily Keywords:}\ }
   {\end{trivlist}}

\journal{Elsvier, License CC-BY-NC-ND 4.0}
\title{System Analysis  and Test-bed for an Atmosphere-Breathing Electric Propulsion System using an Inductive Plasma Thruster}

\author[irs]{F.~Romano\corref{cor1}\fnref{fn1}}
\ead{romano@irs.uni-stuttgart.de}

\author[irs]{B.~Massuti-Ballester\fnref{fn2}}
\ead{massuti@irs.uni-stuttgart.de}

\author[irs]{T.~Binder\fnref{fn3}}
\ead{binder@irs.uni-stuttgart.de}

\author[irs]{G.~Herdrich\fnref{fn4}}
\ead{herdrich@irs.uni-stuttgart.de}

\author[irs]{S.~Fasoulas\fnref{fn5}}
\ead{fasoulas@irs.uni-stuttgart.de}

\author[esa]{and T.~Schönherr\fnref{fn6}}
\ead{tony.schoenherr@esa.int}

\fntext[fn1]{Ph.D.~Student, romano@irs.uni-stuttgart.de}
\fntext[fn2]{Associate Researcher, massuti@irs.uni-stuttgart.de}
\fntext[fn3]{Ph.D.~Student, binder@irs.uni-stuttgart.de}
\fntext[fn4]{Head Plasma Wind Tunnels and Electric Propulsion, herdrich@irs.uni-stuttgart.de}
\fntext[fn5]{Head Department of Space Transportation, fasoulas@irs.uni-stuttgart.de}
\fntext[fn6]{Research Fellow, tony.schoenherr@esa.int}
\address[irs]{Institute of Space Systems (IRS), Universität Stuttgart, Stuttgart, 70569, Germany}
\address[esa]{ESA/ESTEC, Keplerlaan 1, NL-2201 AZ Noordwijk, The Netherlands}

\begin{document}

\begin{abstract}
Challenging space mission scenarios include those in low altitude orbits, where the atmosphere creates significant drag to the S/C and forces their orbit to an early decay. For drag compensation, propulsion systems are needed, requiring propellant to be carried on-board. An atmosphere-breathing electric propulsion system (ABEP) ingests the residual atmosphere particles through an intake and uses them as propellant for an electric thruster. Theoretically applicable to any planet with atmosphere, the system might allow to orbit for unlimited time without carrying propellant. A new range of altitudes for continuous operation would become accessible, enabling new scientific missions while reducing costs. Preliminary studies have shown that the collectible propellant flow for an ion thruster (in LEO) might not be enough, and that electrode erosion due to aggressive gases, such as atomic oxygen, will limit the thruster lifetime. In this paper an inductive plasma thruster (IPT) is considered for the ABEP system. The starting point is a small scale inductively heated plasma generator IPG6-S. These devices are electrodeless and have already shown high electric-to-thermal coupling efficiencies using \ce{O2} and \ce{CO2}. The system analysis is integrated with IPG6-S tests to assess mean mass-specific energies of the plasma plume and estimate exhaust velocities.

\end{abstract}
 \maketitle
\begin{keywords}
ABEP - IPT - IPG - RAM-EP - VLEO
\end{keywords}

\section*{Nomenclature}
\noindent
BOL: Begin-of-life\\
EOL: End-of-life\\
EP: Electric Propulsion\\
FMF: Free Molecular Flow\\
GIE: Gridded Ion Engine\\
HET: Hall Effect Thruster\\
IPG: Inductively Heated Plasma Generator\\
IPT: Inductive Plasma Thruster\\
LEO: Low-Earth Orbit\\
RIT: Radio-Frequency Ion Thruster\\
SA: Solar Array\\
S/C: Spacecraft\\
SSO: Sun-Synchronous Orbit\\
TCS: Thermal Control Subsystem\\
%TIG: Thermionic Generator\\
VLEO: Very Low-Earth Orbit\\
 
\section{Introduction} 
\begin{figure}[h]
	\centering
	\includegraphics[width=9cm]{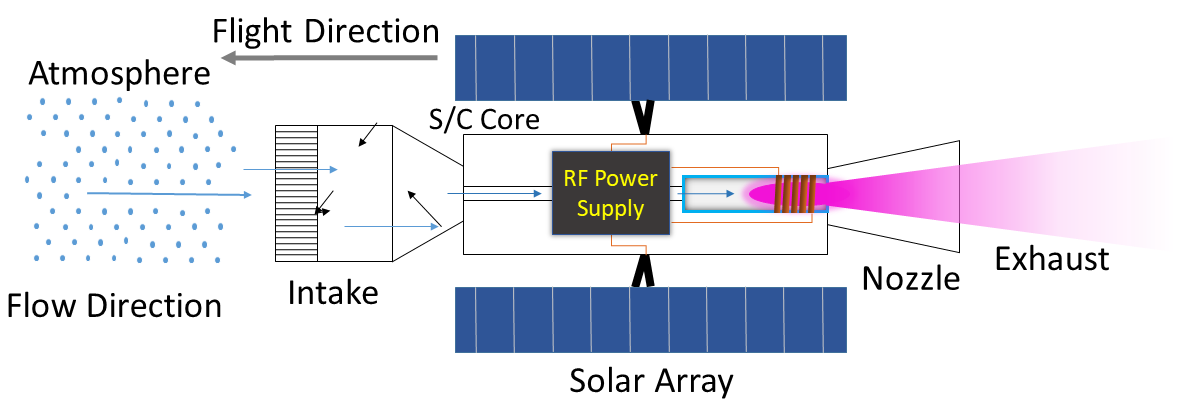}
	\caption{ABEP Concept}
	\label{fig:ABEP}
	\vspace{-20pt}
\end{figure}

Missions in LEO are of great importance for weather forecasting, monitoring of oceanic currents, polar ice caps, fires, agriculture, and military and civil surveillance services. In November 2013 ESA mission GOCE has ended, providing detailed information of Earth geomagnetic field by orbiting as low as~\SI{229}{\kilo\meter}~\cite{GOCE} using QinetiQ T5 gridded ion engines (GIE) as EP. Missions at low altitudes are limited in mission lifetime due to aerodynamic drag, caused by momentum exchange between the residual atmosphere particles and the S/C, requiring an efficient propulsion system that compensates the drag. Such low altitudes would allow simpler and smaller platforms, meaning lower costs, as well as ensuring self de-orbiting at the end of the mission~\cite{llop2014very}. For such missions, the maximum mission lifetime is a mission design driver that depends on the amount of drag that the propulsion system can compensate, and, second, for how long. These two are dependent on the propulsion system efficiency, on the amount of propellant carried on board, and on the generated drag. The basic idea of an Atmosphere-Breathing Electric Propulsion System (ABEP) is to capture the residual atmosphere of the planet, and use it as propellant for an electric thruster, see the concept in Fig.~\ref{fig:ABEP}. This system would ideally nullify the on board propellant required and provide drag compensation, finally increasing mission lifetime. In this paper, a system analysis for an ABEP using an inductive plasma thruster (IPT) as EP is proposed. IPT are electrode-less devices based on inductively heated plasma generators (IPG), therefore eliminating the performance degradation issue typical of GIE and HET, allowing a wide range of propellant to be used and, moreover, removing the need of a neutralizer. IRS has gathered several decades of experience in the development, operation, characterization and qualification of various plasma sources. Among them are steady state self-field and applied field magnetoplasmadynamic (MPD) sources, thermal arcjet devices, IPG and hybrid plasma systems \cite{georg1}, \cite{georg2}, \cite{georg3}. These plasma systems are in application for aerothermodynamic testing, heat shield material characterization \cite{georg4}, \cite{georg5}, \cite{georg6}, \cite{georg7}, \cite{georg8}, \cite{georgacta}, electric space propulsion \cite{georg9}, \cite{georg10}, \cite{georg11}, \cite{georg12}, \cite{georg13}, \cite{georg15}, \cite{georg16} and terrestrial plasma technology (i.e. technology transfer) \cite{georg17}, \cite{georg18}, \cite{georg19}. IPG have originally been developed to cope with chemically aggressive working gases for the IRS plasma wind tunnel PWK3. The electrodeless design enables additionally a pure plasma which engages the potential for aerothermochemical investigations in the field of heat shield material catalysis  \cite{georg6}, \cite{georg8}, \cite{georg20}, nitridation and oxidation \cite{georg21}, \cite{georg22} and, in addition, the behaviour of both plasma sources for plasma wind tunnels and electric propulsion and respective flow conditions \cite{georg23}. Moreover, the high-power inductively heated plasma sources developed at IRS were respectively characterized and modeled to provide increased understanding and an experimental database \cite{georg2}, \cite{georg3}, \cite{georg26}. On basis of both system and mission analyses and the IPG-heritage,  IPG6-S has been tested as IPT candidate in the context of ABEP \cite{6945885}, \cite{myDresden}, \cite{IEPC2015}, \cite{myRome}, \cite{TilmanCanada}. Thrust has been estimated through the measurement of the bulk plasma energy by a cavity calorimeter and compared to the drag derived from the system analysis at the corresponding altitude.
 
\subsection{Literature Review}
A literature review of the most relevant ABEP studies is hereby briefly presented. 
 
ESA~\cite{di2007ram} proposed a technology demonstration mission featuring ABEP. It considers a \SI{1000}{\kilo\gram} S/C equipped with 4 $\times$ RIT-10 operating with the incoming atmosphere as propellant. The S/C is to be set into a circular Sun-Synchronous Orbit (SSO) at an altitude of $h=\SI{200}{\kilo\meter}$ for a 7 years mission. Front area is of $\SI{1}{\meter^2}$ and the maximum power available for propulsion is of $\SI{1}{\kilo\watt}$, enabling thrust from $2$ to \SI{20}{\milli\newton}. Solar array (SA) surface is of $\SI{19.74}{\square\meter}$ generating a EoL power of $\SI{2.9}{\kilo\watt}$ combined with a \SI{612}{\watt\hour} \ce{Li}-Ion battery.

Diamant~\cite{dia} proposes a mission for a small S/C with drag compensation at $h=\SI{200}{\kilo\meter}$ with a 2-stage cylindrical Hall thruster and propellant ingested from the atmosphere. The first stage is an electron cyclotron resonance (ECR) ionization stage and the second stage is a cylindrical HET. The required power is of $\SI{1}{\kilo\watt}$ for propulsion, the frontal area is of $\SI{0.5}{\meter^2}$ with a collection efficiency $\eta_c$, ratio between amount of encountered atmosphere particles and that of delivered to the thruster, of $35\%$.
 
Shabshelowitz~\cite{shabshelowitz2013study} investigates RF Plasma applied to an ABEP system. The S/C mass is of $\SI{325}{\kilo\gram}$, to be set into a circular orbit at an altitude of $\SI{200}{\kilo\meter}$ for a mission duration of 3 years. The frontal area is of $\SI{0.39}{\meter^2}$, a length of $\SI{2.1}{\meter}$, and the S/C has a cylindrical shape to be covered with solar cells. The ratio of the frontal area through the inlet area is of $A_{f}/A_{inlet}=0.5$ and $\eta_c$ is assumed of $90\%$. The propulsion system is composed by a single-stage HET operating with air and supported by a tank of propellant for ballast. The thruster requires a power of $\SI{306}{\watt}$.
 
Pekker and Keidar~\cite{pekker2012analysis} considered a HET using the atmosphere as propellant, orbiting at $h=\SI{90}{\kilo\meter}$ and $h=\SI{95}{\kilo\meter}$. Gas leaving the chamber of the HET is considered fully ionized and the estimated achievable thrust is of $F_{T @  \SI{90}{\kilo\meter}}=\SI{22}{\newton}$ and of $F_{T @ \SI{95}{\kilo\meter}}=\SI{9.1}{\newton}$ with a thrust density of $\SI{13}{\milli\newton\per{\kilo\watt}}$. Power required at the two altitudes is of $P_{req @ \SI{90}{\kilo\meter}} = 1.6-\SI{2}{\mega\watt}$ and $P_{req @ \SI{95}{\kilo\meter}}=700-\SI{800}{\kilo\watt}$.

The ESA GOCE mission successfully ended in November 2013. The S/C had a mass of $\SI{1050}{\kilo\gram}$ and orbited into a $250$ (finally $235$)$ - \SI{265}{\kilo\meter}$ SSO for a predicted mission lifetime of  $20-30$ months, but it reached finally 4 years of operation. The frontal area was of $\SI{1.1}{\square\meter}$~\cite{GOCE}. The S/C was provided with two GIE or Kaufman-type ion engines, derived from the QinetiQ T5 (one for backup) operating with \ce{Xe} and providing thrust between $T = 1.5$ and $\SI{20}{\milli\newton}$. The SA for the power subsystems provided $P_{EOL}=\SI{1.6}{\kilo\watt}$ in EOL suported by $\SI{78}{\ampere\hour}$ battery.

The BUSEK company~\cite{busek} developed an ABEP concept applied to a small S/C orbiting Mars: Martian Atmosphere-Breathing Hall Effect Thruster (MABHET). An HET has been operated with a gas mixture reproducing Mars' atmosphere, mostly dominated by \ce{CO2}. The thrust to power peak ratio of HET has been measured to be \SI{30}{\milli\newton\per{\kilo\watt}} with a low peak of \SI{19}{\milli\newton\per{\kilo\watt}}. The inlet area is of $\SI{0.15}{\square\meter}$ and the frontal area $\SI{0.30}{\square\meter}$. The collection efficiency is of $\eta_c=35\%$. Compression of the incoming air flow is required to achieve better performance of the thruster. MABHET may work better in Mars than in Earth's orbit because of lower density and temperature of the atmosphere and, moreover, different accommodation coefficients.
 
The study from JAXA~\cite{JAXA} is a concept for an Air-Breathing Ion Engine (ABIE). Atmospheric propellant is ionized by an ECR device. A S/C has been proposed orbiting in a circular polar SSO of $h=\SI{170}{\kilo\meter}$ for at least 2 years. The frontal area is of $\SI{1.5}{\square\meter}$ with an inlet area of $\SI{0.48}{\square\meter}$ providing an intake efficiency up to $\eta_c=46\%$. The propulsion system should deliver a thrust to power ratio between $10-\SI{14}{\milli\newton\per{\kilo\watt}}$. Moreover, altitudes of $180$ and \SI{140}{\kilo\meter} have been investigated, requiring a power for the thruster of \SI{470}{\watt} and \SI{3.3}{\kilo\watt}.
A summary of the literature review is briefly shown in Tab.~\ref{tab:lite}.\begin{table}[h]
 \caption{Summary of Literature Review Results}
 \centering
 \begin{tabular}{ll}
 \toprule
 Quantity & Value\\
 \midrule
 S/C Mass &$<\SI{1050}{\kilo\gram}$\\
 Inlet Area & $0.3-\SI{1}{\square\meter}$\\
 Orbit & SSO, $90-\SI{250}{\kilo\meter}$\\
 Mission Duration & $2-8$ years\\
 Thrust Density & $10-\SI{59}{\milli\newton\per{\kilo\watt}}$\\
 Power & $0.660-\SI{3.3}{\kilo\watt}$\\
 Collection Efficiency & $0.35-0.9$\\
 \bottomrule
 \end{tabular}
 \label{tab:lite}
 \end{table}

Based on this literature review, the candidate IPT thruster, IPG6-S, is collocated in the power range $<\SI{3.5}{\kilo\watt}$, with a discharge channel internal diameter of \SI{37}{\milli\meter}, using a reference frontal area $A_f$ equal to that of the intake area $A_{in}$ of \SI{1}{\square\meter} for the estimation of the drag and the collectible mass flow, assuming literature values of $\eta_c=0.35-0.9$. The mission altitude, for drag and mass flow calculation, is below \SI{250}{\kilo\meter} in Earth orbit.
 
\section{System Analysis}
In this section the system analysis for an ABEP mission is presented. It takes into account the atmospheric model used to extrapolate the input required for the estimation of collectible mass flow and generated drag. Considerations regarding the target orbit and basics for the intake design are included. Calculation of the required SA is also performed. 
\subsection{Atmospheric Model}
\begin{figure}[h]
   %\centering
   \includegraphics[width=13cm]{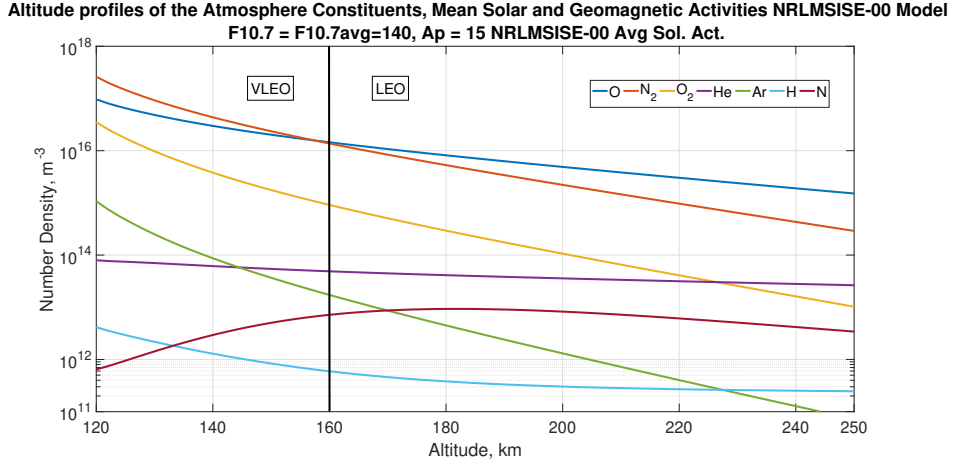}
   \caption{Earth's atmosphere composition.}
   \label{fig:model}  
   \vspace{-10pt}  
\end{figure}

\begin{figure}[h]
      \centering
      \includegraphics[width=13cm]{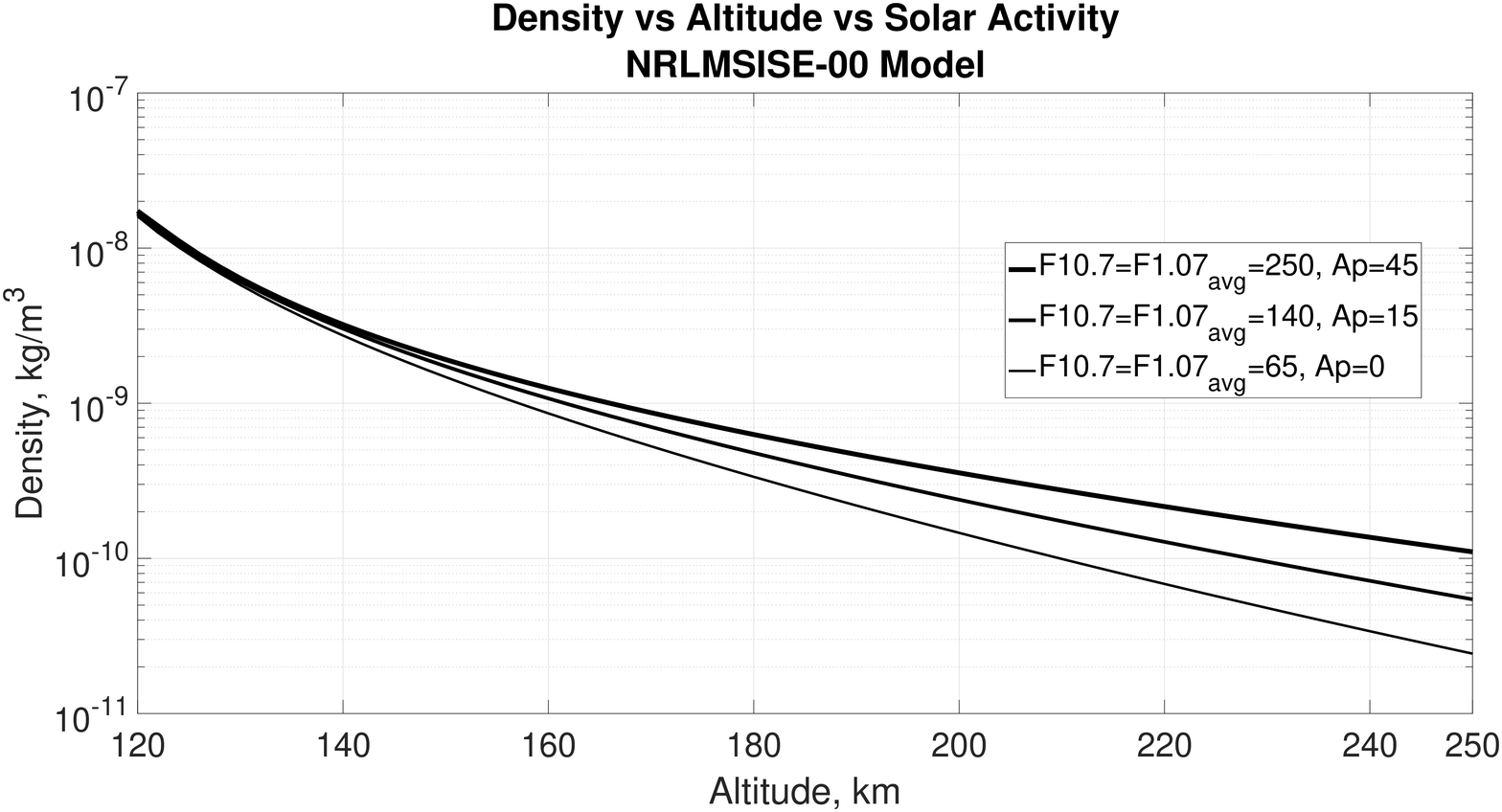}
      \caption{Density vs. altitude and solar activity.}
      \label{fig:model1}    
      \vspace{-10pt}
      \end{figure}

The chosen atmospheric model for Earth is the NRLMSISE-00. Compared to MSISE-90, it provides better estimation of the atmosphere density below $h=\SI{350}{\kilo\meter}$ and it is the most accurate model for residual atmosphere composition in LEO and VLEO~\cite{picone}. It is an empirical global model of Earth's atmosphere under different conditions of solar and geomagnetic activities. The data have been generated through the NRLMSISE-00 model website. Inputs are date, geographical coordinates and solar activity parameters, $F10.7$ and $Ap$. Results show that the most dominant elements in VLEO and LEO are \ce{O2} and \ce{N2}, with the first more dominant at higher altitudes, as shown in Fig.~\ref{fig:model}. Particular care must be taken concerning the solar activity which cycles every 11 years. This will result in change of the density vs.~altitude profile as it compresses and releases the atmosphere by time, as shown in Fig.~\ref{fig:model1}, the change is greater for a higher altitude in VLEO range.  Moreover, variation of density arise also due to different location over the Earth, as well as day/night density variations in a non-SSO orbit. The number densities of the species also changes over time, position, and altitude. These variations will affect the thruster performance and also the drag. Therefore, the thruster requires a certain flexibility in terms of propellant and performance, to cope with the different densities and atmosphere composition. 

 \subsection{Orbit}
 LEO extends in the range from 160 to \SI{2000}{\kilo\meter}, VLEO from 100 to \SI{160}{\kilo\meter}~\cite{6945885}. According to ESA~\cite{di2007ram} the upper altitude limit for an ABEP mission is \SI{250}{\kilo\meter} to be competitive against conventional EP. The minimum altitude has been set according to JPL,~\cite{you}, at \SI{120}{\kilo\meter}, due to heating effects, moreover, this altitude is considered as that at which re-entry is engaged~\cite{bilbey2005investigation}. Concerning the orbit plane, considering to continuously generate power with SA, a SSO is chosen. In an SSO the sun vector is always perpendicular to the orbit plane, therefore directing SA in the orbit plane will theoretically enable to operate at maximum power for most of the time. However, this depends on the mission requirements. If a particular orbit is required, depending on the propulsion system requirements in terms of electrical power, a thrust profile related to the eclipse and sunshine periods has to be investigated.
 
\subsection{Intake}
\begin{wrapfigure}{r}{.4\textwidth}
	\vspace{-25pt}
	\begin{center}
		\includegraphics[height=4cm]{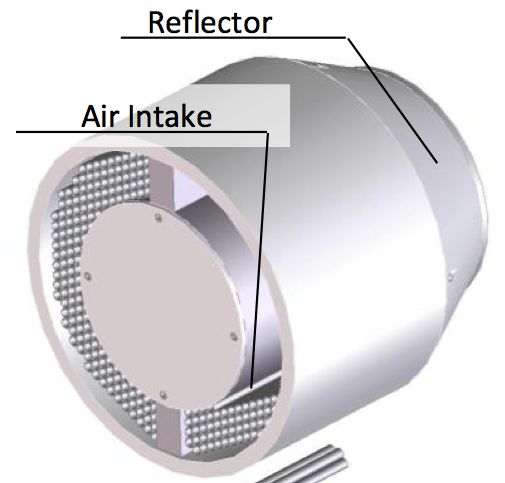}
	\end{center}
	\vspace{-10pt}
	\caption{JAXA Concept~\cite{JAXA}.}
	\label{fig:jaxaint}    
	\vspace{-10pt}
\end{wrapfigure}
\begin{figure}
    \centering
    \includegraphics[width=13cm]{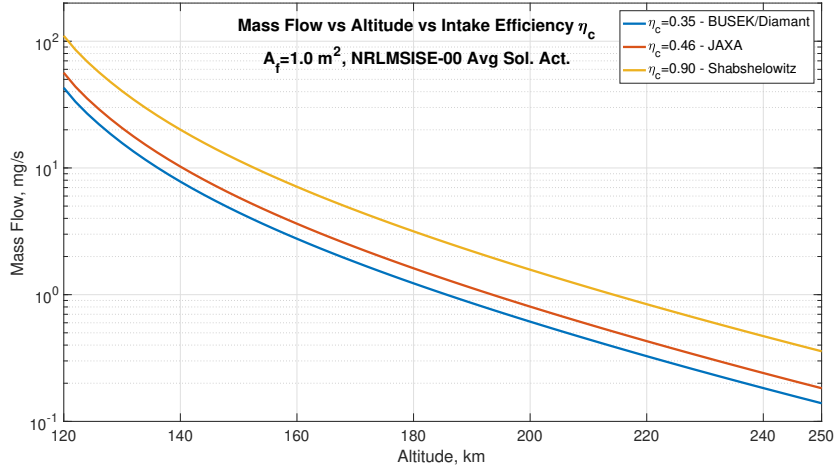}
    \caption{Mass flow vs. altitude and $\eta_c$ for a fixed $A_f$.}
    \label{fig:mdot}    
    \end{figure}

 The intake is the device that collects and delivers the atmosphere particles to the thruster.According to~\cite{6945885} a mechanical device should be used as the ionization degree in LEO and VLEO is too low to use a magnetic one. JAXA~\cite{JAXA} developed an ABEP system that combines intake and thruster into one apparatus, see Fig.~\ref{fig:jaxaint}. This design also embodies some of the S/C subsystems in a concentric cylinder inside the intake, while the atmosphere particles are collected in the outer ring region. A collection efficiency up to $40\%$ and a compression factor between $100-200$~\cite{JAXA2} are achievable and have been verified~\cite{IEPC2015}, a pressure of \SI{1}{\milli\pascal} is achieved at the thruster head but is not enough for most conventional EP~\cite{6945885}. A different approach is that of BUSEK~\cite{busek}, a long open duct with an end cone, to allow more compression due to multiple collisions cascade phenomena. Both of these two designs includes an inlet structure of small ducts to reduce the backflow escaping from the intake, see Fig.~\ref{fig:jaxaint}. The incoming flow has high velocity and is collimated, thus, the probability of passing through small ducts with little interaction is high. As they reach the end of the intake they are reflected back in a random direction and they loose most of their velocity, therefore the probability of flowing back to space through the inlet structure is smaller~\cite{IEPC2015}. In Fig.~\ref{fig:mdot}, mass flow vs. altitude is plotted for average solar activity considering an intake area of $A_{in}=A_f=\SI{1}{\square\meter}$ and $\eta_c$ from the literature review: $\eta_c=0.35$~\cite{dia} and~\cite{busek},~$0.46$~\cite{JAXA},~$0.90$~\cite{shabshelowitz2013study}.
In our recent studies~\cite{IEPC2015},~\cite{myRome},~\cite{TilmanCanada}, optimization has been done showing a strong relation between $A_{in}$ and $A_{thr}$, the thruster cross section, regarding how much mass flow can be collected and how efficiently. This generates a loop for the design between $A_{in}/A_{thr}$, $\dot{m}_{thr}$, and $\eta_c$, where the last two cannot be maximized at the same time. For the scope of this paper, the three fore-mentioned $\eta_c$ will be used as they remain within the range of the optimized values calculated in our recent studies~\cite{IEPC2015},~\cite{myRome}.
    
\subsection{Drag}
\begin{figure}
        \center
        \includegraphics[width=13.5cm]{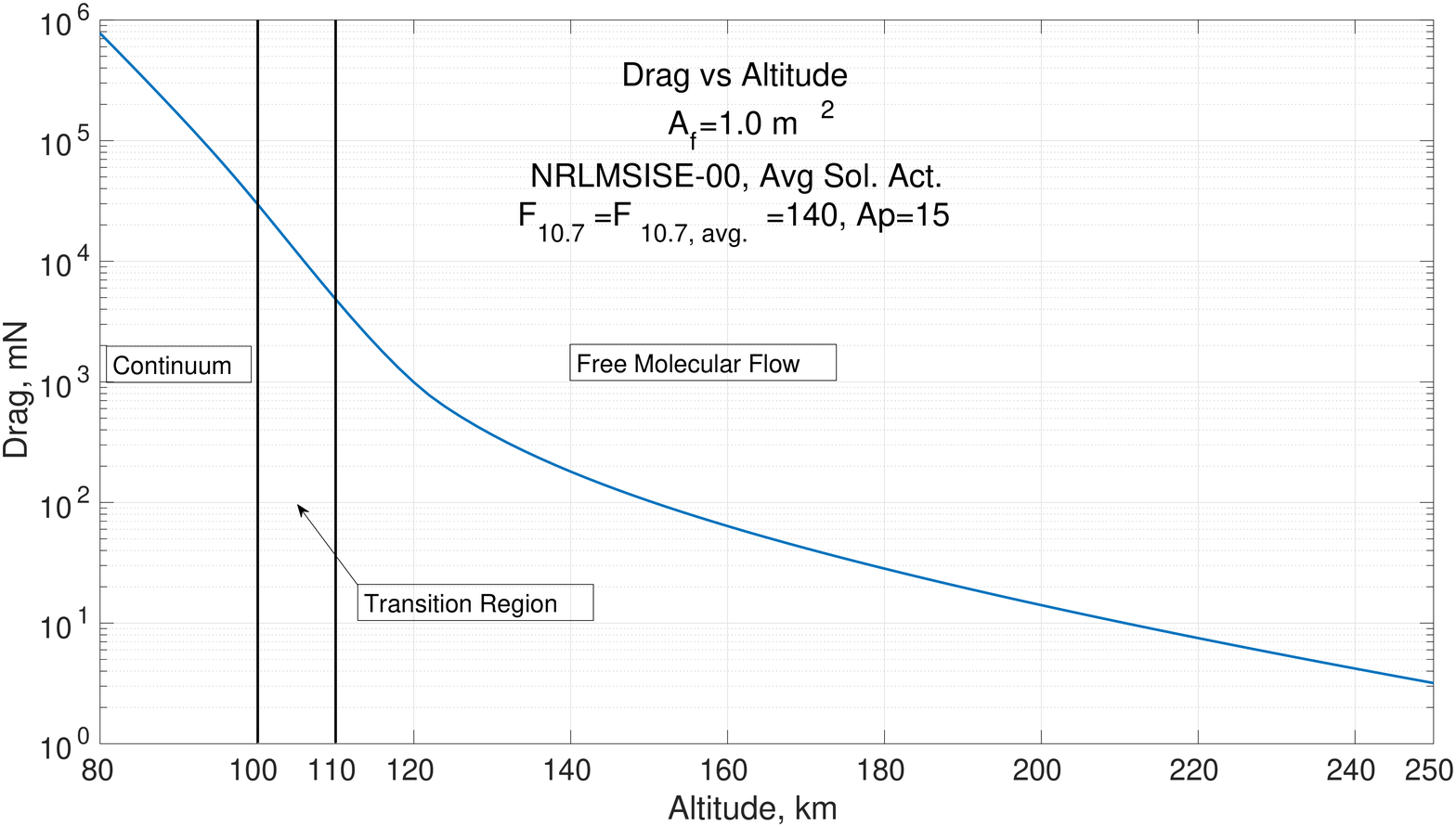}
        \caption{Drag vs. altitude, Earth.}
        \label{fig:drag}   
        \vspace{-10pt} 
        \end{figure}

\begin{figure}
        \center
        \includegraphics[width=13cm]{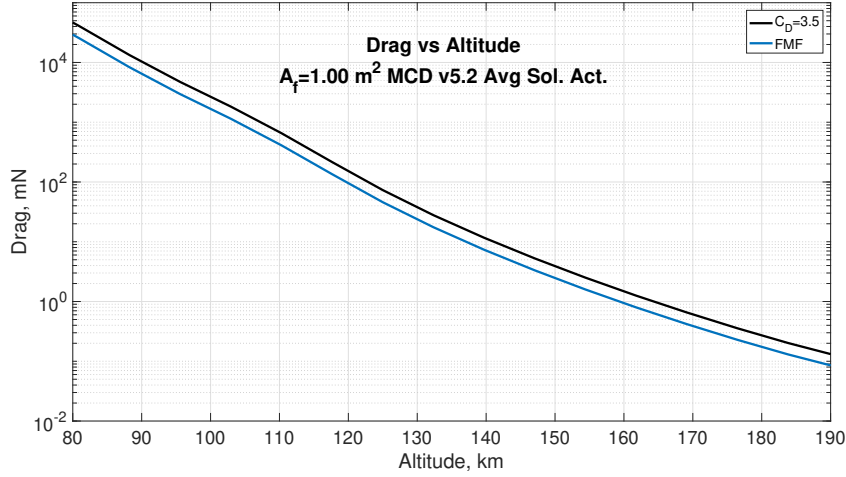}
        \caption{Drag vs. altitude, Mars.}
        \label{fig:drag_Mars}    
        \end{figure}

Drag estimation is fundamental for the mission design. S/C will orbit at low altitudes, where the presence of residual atmosphere is not negligible, as it will slow down the S/C. Due to the particular conditions at these altitudes, the first step is to determine which kind of flow the S/C is flying into. Whether the flow is to be considered continuum or free molecular (FMF), where the mean free path length between molecules~$\lambda$ becomes comparable to the S/C size~$L$ and particle collisions can be neglected, is determined by the Knudsen number $Kn=\lambda/L$. Estimations show that for altitudes above \SI{120}{\kilo\meter} in Earth orbit, and mean length of $L=0.3,~1,~2,~\SI{3}{\meter}$, the flow is FMF. A sensitivity analysis on the average molecules size and on the solar activity has been done and shows very little variations. The effect of solar activity in Earth orbit is more appreciable on higher altitudes rather than lower, see Fig.~\ref{fig:model1}.\\
For the calculation of the drag, the drag equation in Eq.~\ref{eq:freeflow} has been implemented. A drag coefficient of $C_D=2.2$ has been exemplary selected, which is a typical average in literature for small S/C in LEO, and for normal facing area according to~\cite{shen}. To increase the collection efficiency, the intake might take advantage from specular reflections, therefore reducing $C_D$ to even smaller values. The use of specular reflecting surfaces, might also allow to reduce overall produced drag of the spacecraft. However, a worst condition is chosen assuming full accommodation of the particles at the front and intake surfaces. Future work includes the estimation of $C_D$ due to the intake in front of the spacecraft.\begin{equation}
    F_D=\frac{1}{2}\rho(h)A_f v_{rel}^2 C_D
    \label{eq:freeflow}
    \end{equation}
$F_D$ is the drag force, $\rho(h)$ is the atmosphere density, $v_{rel}$ is the velocity of the S/C relative to the atmosphere, $A_f$ is the front surface facing the flow. From this preliminary drag estimation, the contribution of solar arrays is neglected by considering them to be parallel to the direction of flight, therefore creating a minimum contribution to the frontal area, such a configuration can be found in the GOCE spacecraft~\cite{GOCE}. Lateral surfaces also have impact on the drag, however, these require DSMC simulation and will be evaluated in further work . The result of this calculation is shown in Fig.~\ref{fig:drag}. In particular the transition region is defined where the $Kn$ variates from $0.1$ (continuum flow) to 1 (FMF), this is in the altitude range between $100$ and \SI{110}{\kilo\meter} for $L=\SI{1}{\meter}$.
This result has been compared to the statistical model according to~\cite{shen} and had shown good match with Eq.~\ref{eq:freeflow} and the corresponding $C_D$. The model is based on Eq.~\ref{eq:shen1},~\ref{eq:shen2},~\ref{eq:shen3}, and~\ref{eq:shen4}.
\begin{equation}
p=\frac{1}{2} \rho {\frac{v_{rel}^2}{S^2}}(X_w+Y_w)
\label{eq:shen1}
\end{equation}
\begin{equation}
S=\frac{v_{rel}}{\sqrt{2R T_a}}
\label{eq:shen2}
\end{equation}
\begin{equation}
X_w=\biggl( \frac{2-\sigma}{\sqrt{\pi}}S\cos{\theta}+\frac{1}{2}\sigma \sqrt{\frac{T_r}{T_a}}\biggr)e^{-(S\cos{\Theta})^2}
\label{eq:shen3}
\end{equation}
\begin{equation}
Y_w=\biggl[(2-\sigma)\biggl[(S\cos{\theta})^2+0.5\biggr]+\frac{1}{2}\sigma\sqrt{\pi\frac{T_r}{T_a}}S\cos{\theta}\biggr](1+erf(S\cos{\theta}))
\label{eq:shen4}
\end{equation}
The pressure on the assumed flat plate is $p$, $S$ is the molecular speed ratio, $\sigma$ is the accommodation coefficient set to 1 for complete diffusive accommodation, $T_r$ temperature of the reflected particles is set to that of the S/C at $\SI{300}{\kelvin}$, $T_a$ is that of the atmosphere, $\theta$ is the angle of attack set to \SI{0}{\degree} and \textit{erf} is the error function. Finally, the drag is calculated as Eq.~\ref{eq:shen5}.
\begin{equation}
F_D=pA_f
\label{eq:shen5}
\end{equation}
The model considers particle-surface interactions, S/C temperature, reflected particles temperature. However, these additional variables imply more assumptions on an already simplified model, therefore the simply approach from Eq.~\ref{eq:freeflow} has been chosen. Care must be taken in the lowest altitudes, $h<\SI{120}{\kilo\meter}$ for Earth, as assessing of heating rates is necessary. Drag has also been estimated for a S/C orbiting around Mars, the result is shown in Fig.~\ref{fig:drag_Mars} for both Eq.~\ref{eq:freeflow} and Eq.~\ref{eq:shen5}.

\subsection{Power Supply}
The power supply subsystem must provide electrical power for the propulsion system together with all the other subsystems. A common approach for S/Cs orbiting Earth or Mars is the use of solar arrays (SA) and batteries. Batteries compensate the fluctuation of required power and provide electricity when the SA are not illuminated by the Sun. SA provide electric power to all the subsystems and recharge the batteries when illuminated by the Sun.\\
%The physical quantity that becomes important when orbiting at very low altitudes is the heat produced by the impact of the residual atmosphere particles. At very low altitudes this heat has to be dissipated to avoid S/C overheating. %This can be also converted directly into electricity by the use of a thermionic generator (TIG) that operates on the principle of the Seebeck effect, which describes the phenomena of voltage generation in a conductor or semiconductor when subjected to a temperature gradient. Thermionic generators have no moving parts, but they are not yet a mature technology. A recent study, see~\cite{thermionic}, calculated a maximum efficiency of $\eta = 42\%$ in optimum conditions.\\
%This technology might reduce the required S/A total surface, resulting in less drag and an efficient use of the heat. It has to be kept in mind that only a fraction of the heat can be converted into electrical energy, the remaining part must be still dissipated by the TCS. 
SA with a minimum average BOL efficiency of $\eta = 29.5\%$ have been considered,~\cite{sa}. Sun vector has been assumed always perpendicular to the SA, as in an SSO. The calculated required areas are in Tab.~\ref{tab:SA1} and \ref{tab:SA2} for both power and voltage considering the SA degradation over a 7 years long mission, according to the ESA study~\cite{di2007ram}.
\begin{table}[h]
\parbox{.40\linewidth}{
\caption{Power vs. SA Area, EOL}
\label{tab:SAP}
\centering
\begin{tabular}{ll}
\toprule
$P_{max}$ & $A_{SA}$\\
{\SI{}{\kilo\watt}} & {\SI{}{\square\meter}}\\
\midrule
0.5 & 2.0\\
1 & 4.0\\
1.5 & 6.0\\
%2 & 7.8\\
3 & 11.9\\
3.5 & 13.8\\
5 & 19.6\\
\bottomrule
\end{tabular}
\label{tab:SA1}
}
%hfill
\parbox{.60\linewidth}{
\vspace{-35pt}
	\caption{Voltage vs. String Area - EOL}
	\label{tab:SAV}
	\centering
	\begin{tabular}{ccc}
	\toprule
	Voltage & No. of Cells & $A_{string}$\\
	{\SI{}{\volt}} & - &  {\SI{}{\square\meter}}\\
	\midrule
	550 & 319 & 0.85\\
	850 & 493 & 1.30\\
	1000 & 579 & 1.54\\
	\bottomrule
	\end{tabular}
\label{tab:SA2}
}
	\end{table}

\section{IPG6-S, Experimental Set-Up}
The candidate test-bed ABEP thruster is an inductively heated plasma generator (IPG), more precisely IPG6-S available at IRS facilities is used. The IPG main advantage is the electrode-less design. No critical components have direct contact with the plasma, therefore any issues concerning erosion are eliminated. In addition, the IPG does not employ a grid system that could suffer from erosion. Moreover, the IPG functional principle enables ignition also at very low densities. The advantages as propulsion system, are as following:\begin{itemize}
	\item Electrode-less design, no accelerating grids;
	\item No neutralizer needed;
	\item Less sensitive in terms of minimum pressure and mass flow for ignition.
\end{itemize}
Indeed, the high presence of \ce{O} and \ce{O2} in LEO and VLEO, main erosion responsible on grids/electrodes/discharge channels, will decrease thruster performance over time, such that of RIT, erosion of the accelerating grids, or of HET, where the discharge channel is eroded. In addition, plasma leaving the IPG is already neutral, eliminating the need of a neutralizer.

Experiences with the high power IPG3 in the field of experimental aerothermodynamics and first assessments of IPG7 as a high power electric thruster, show promising results with respect to the propellant flexibility of such devices. Here, propellants such as \ce{O2}, \ce{N2}, \ce{CO2}, water vapour, \ce{Ar}, and blended propellant systems could be successfully qualified~\cite{herdrich2002operational},~\cite{owens2010development}.

\subsection{Facility Description}
\begin{wrapfigure}{r}{.6\textwidth}
	\vspace{-10pt}
	\centering
	\includegraphics[width=6.5cm]{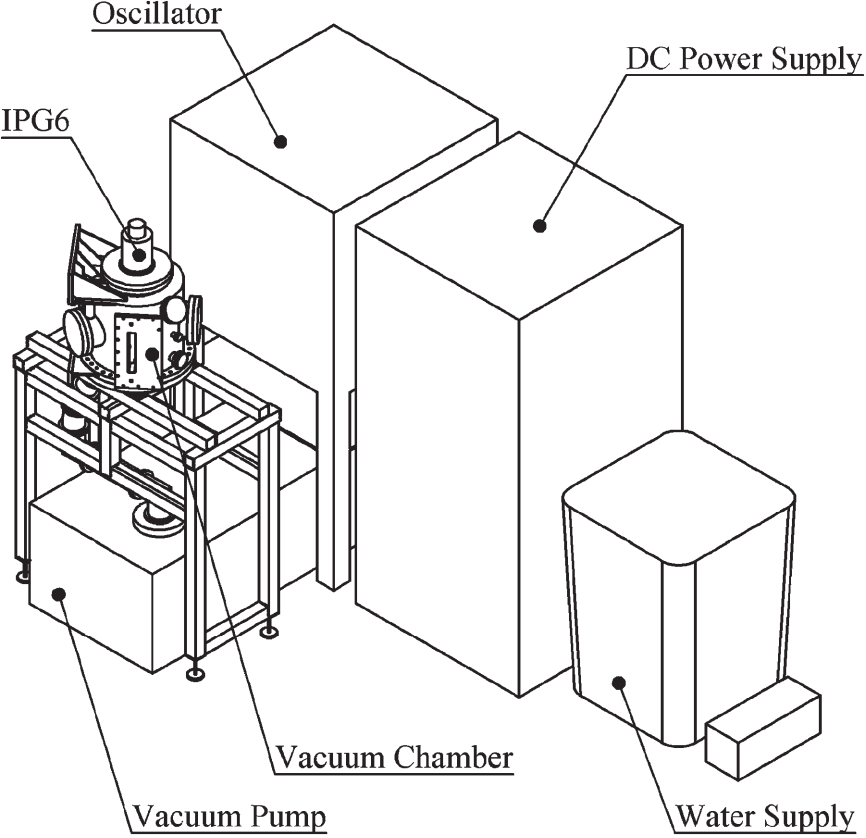}
	\caption{IPG6-S Facility}
	\vspace{-10pt}
	\label{fig:facility}
\end{wrapfigure}
\begin{figure}[h]
	\centering
	\includegraphics[width=9.5cm]{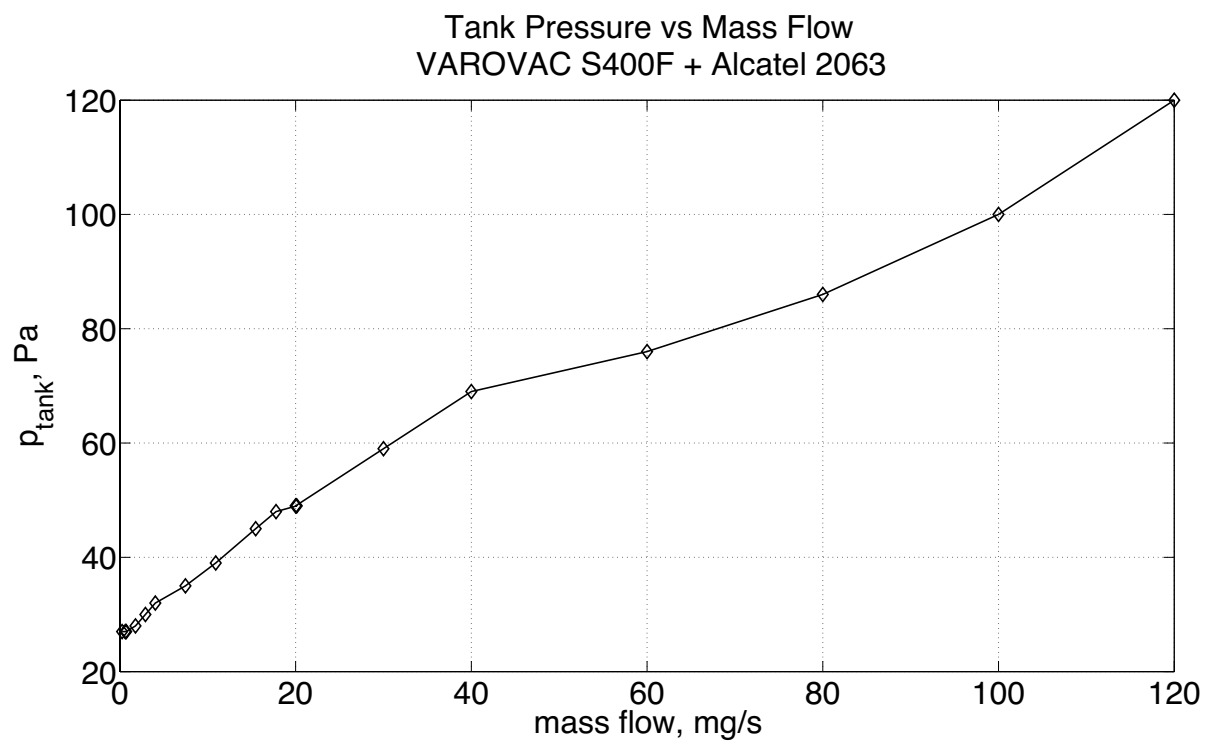}
	\caption{Vacuum Chamber Pressure Dependence on Mass Flow}
	\vspace{-10pt}
	\label{fig:tankpressure}
\end{figure}

The experimental set-up used to run IPG6-S is described within this section. IPG6-S is mounted on top of a \SI{0.4}{\meter} diameter and length vacuum chamber, as shown in Fig.~\ref{fig:facility}. The vacuum system is composed by two vacuum pumps mounted in parallel: a single stage rotary vane Leybold VAROVAC S400F with a nominal pumping speed of \SI{400}{\cubic\meter\per{\hour}}, and a two-stage vane pump Alcatel 2063 with a nominal pumping speed of \SI{65}{\cubic\meter\per{\hour}}. The vacuum pumps have been always operated at the same time due to safety issues. The vacuum chamber base pressure reached \SI{26}{\pascal} without mass flow, the behavior with mass flow is shown within Fig.~\ref{fig:tankpressure}.
The electrical power is provided by the RF generator HGL 20-4B, that provides a maximum of \SI{20}{\kilo\watt} power at a frequency of \SI{4}{\mega\hertz}. It is composed by a DC power supply, and an oscillator that feeds the IPG through RF power lines, as shown in Fig.~\ref{fig:ipg6s}. The DC power supply is cooled by air. The oscillator box is provided with the tetrode and the resonant circuit, including the IPG6-S coil, that are water cooled. Anode current, screen grid voltage, and resulting active power are shown in analog displays at the power supply. The screen grid voltage regulates how much current can flow at the anode of the tetrode. Pressure is measured by two Pfeiffer PKR 251 Full-Range gauges, one for the vacuum chamber, and one for the injector head of IPG6-S. 
Gas supply is provided by two Bronkhorst flow controller F-201AV-50K that provide between 20 and 230$-\SI{250}{\milli\gram\per{\second}}$, and one Tylan FC-2900 for mass flows below \SI{20}{\milli\gram\per{\second}}.
The water cooling subsystem operates in a closed circuit by means of two water pumps, a water tank, and a heat exchanger. Water cools IPG6-S casing and quartz tube, injector, bottom flange, VAROVAC vacuum pump, the tetrode and the resonant circuit (including IPG6-S coil) of the power supply, and the cavity calorimeter.\begin{wrapfigure}{r}{.45\textwidth}
	\vspace{-40pt}
	\centering
	\includegraphics[height=3.5cm]{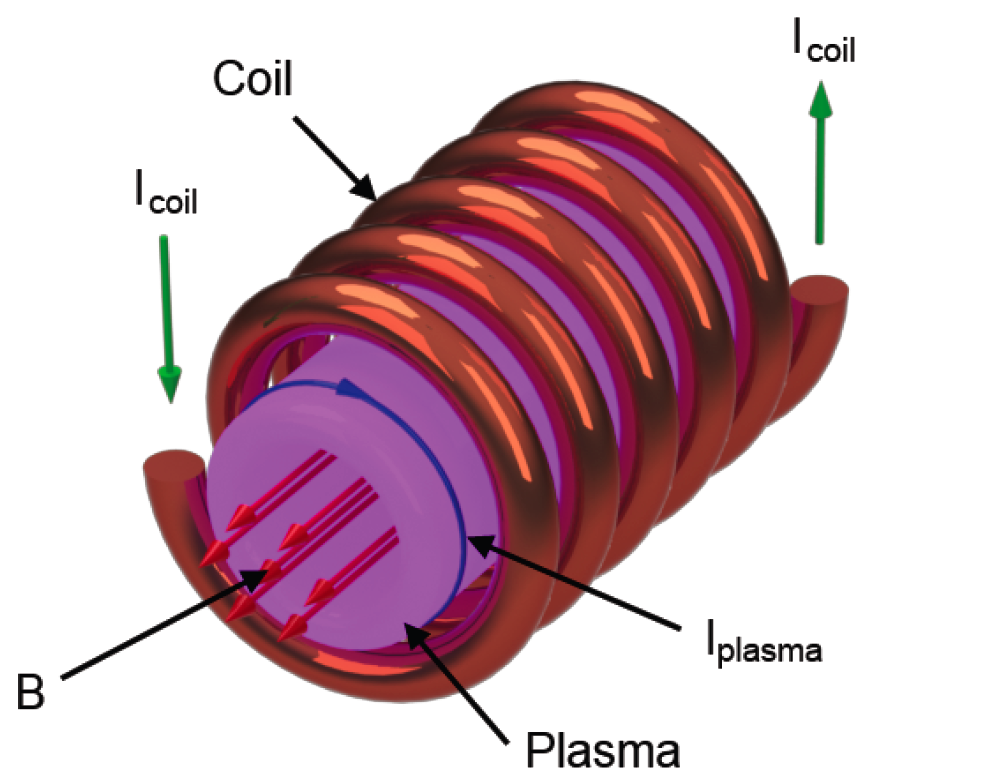}
	\vspace{-5pt}
	\caption{IPG Principle of Operation}
	\label{fig:op_princ}
	\vspace{-10pt}
\end{wrapfigure} 

\subsection{IPG Principle of Operation}
In an IPG a coil is wrapped around a quartz tube, the discharge channel, and fed by RF AC current. It operates in a way similar to a transformer where the primary winding is the coil and the secondary is the gas inside the discharge channel. The current flowing in the coil induces an oscillating B-field in the discharge channel which accelerates ions and electrons of the gas. The oscillating B-field is modeled by a respectively radial E-field. The E-field is responsible for the plasma current, circulating in the opposite direction to that of the coil current. a chain reaction establishes, increasing both temperature and electrical conductivity of the plasma, see Fig.~\ref{fig:op_princ}. Moreover, there is a likelihood of using a self-field that could create a theta pinch effect~\cite{petkow}. 

\subsection{IPG6-S}
\begin{wrapfigure}{r}{.45\textwidth}
	\vspace{-20pt}
	\centering
	\includegraphics[height=6.5cm]{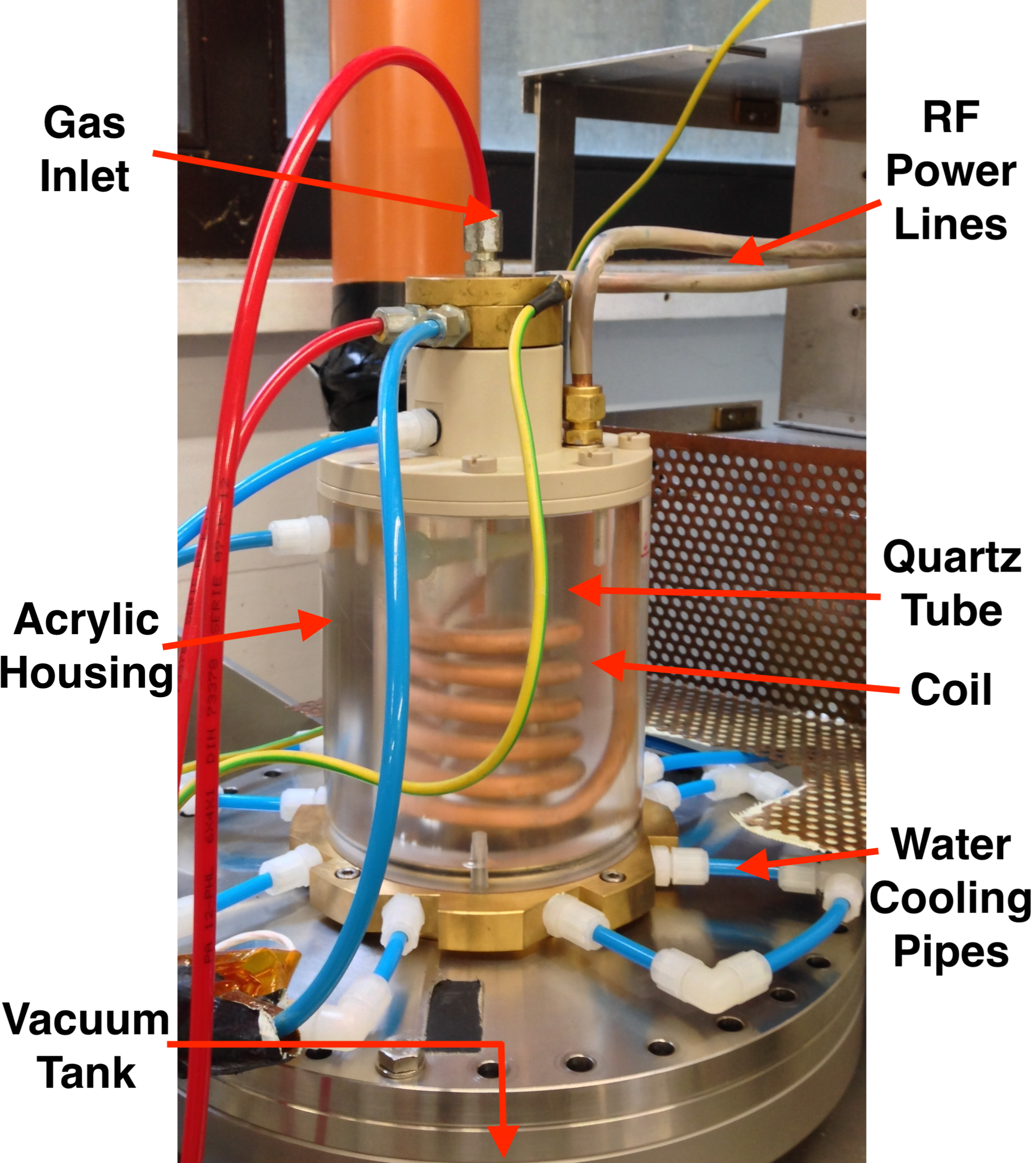}
	\caption{IPG6-S}
	\label{fig:ipg6s}    
\end{wrapfigure} IPG6-S, see Fig.~\ref{fig:ipg6s}, has been used for the tests. It has been chosen for its size and power levels that are scalable to a small S/C~\cite{myDresden}. The power supply provides a maximum input power $P_{max}=\SI{20}{\kilo\watt}$, an anode current up to \SI{4}{\ampere}, and an anode voltage of $7.7, 8.2, \SI{8.5}{\kilo\volt}$. The frequency is of $f\sim\SI{4}{\mega\hertz}$ depending on the impedance of the IPG. During operation the active power never exceeded \SI{3.5}{\kilo\watt}. IPG6-S is water cooled, the discharge channel has an inner diameter of \SI{37}{\milli\meter}, a length of \SI{180}{\milli\meter} and the coil has 5.5 turns providing an inductance of $\SI{0.489}{\micro\henry}$~\cite{tomeu}. A twin facility, IPG6-B, is installed at the University of Baylor, Waco, Texas, USA~\cite{dropmann}.

\subsection{Tests}
\begin{wrapfigure}{r}{.25\textwidth}
	\vspace{-30pt}
	\begin{center}
		\includegraphics[height=6cm]{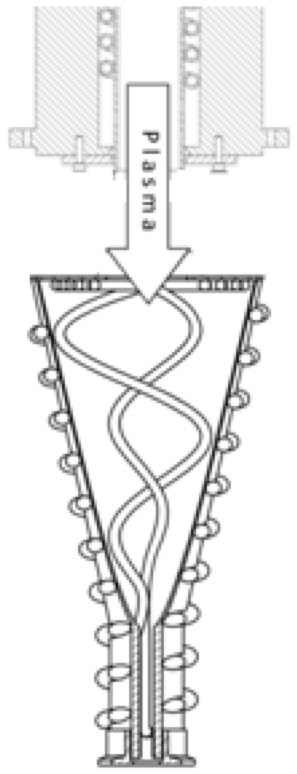}
	\end{center}
	\vspace{-15pt}
	\caption[IPG6-S Calorimeter.]{Cavity Calorimeter~\cite{calorimeter}}
	\label{fig:calorimeter}
	\vspace{-20pt}
\end{wrapfigure}The input required for the test are in terms of propellant, mass flow, and applied screen grid voltage. As result from the system analysis \ce{N2} and \ce{O} are the elements most present in LEO and VLEO. As first approach air-only, as it is composed by $\sim78\%$ of \ce{N2}, to simulate \ce{N2} operation, and \ce{O2}-only, due to the difficulty of storing \ce{O}-only, to simulate atomic oxygen operation have been used, sweeping the mass flow from $~0.2$ to \SI{120}{\milli\gram\per{\second}}. Each mass flow has been assigned to a certain altitude, by comparing the total collectible mass flow from an intake that has a front area of $A_f = \SI{1}{\square\meter}$ and two collection efficiencies of $\eta_c=1$ and $0.35$, all for average solar activity. Three screen grids voltages, that regulate the current flow, have been selected for the tests: $0.55$, $0.85$, and $\SI{1.00}{\kilo\volt}$.

\subsection{Assessment of Thruster Relevant Parameters}
In order to assess thrust relevant parameters for IPG6-S, the following procedure has been performed. A cavity calorimeter is a device, see Fig.~\ref{fig:calorimeter}, used to evaluate plasma energy by measuring the temperature difference of the water, between inlet and outlet of the device, that exchanges heat with the plasma leaving the generator. The estimation is done according to Eq.~\ref{eq:cal}, where $h_{cal}$ is the enthalpy of the plasma, $\dot{m}_{gas}$ is the gas mass flow in the IPG, $P_{cal}$ is the measured calorimeter power, $\dot{m}_{w}$ is the water flow in the calorimeter, $C_{{p}_{w}}$ is the water specific heat capacity at \SI{20}{\degreeCelsius}, and $T_{out,in_{cal}}$ is the water temperature at the outlet and inlet of the calorimeter.\begin{equation}
h_{cal}=\frac{P_{cal}}{\dot{m}_{gas}}=\frac{\dot{m}_{{w}_{cal}} C_{{p}_{w}} (T_{out,cal}-T_{in,cal})}{\dot{m}_{gas}}
\label{eq:cal}
\end{equation}
It is assumed that all the plasma energy measured by the calorimeter is converted into kinetic energy. Therefore the exhaust velocity, $c_e$, can be estimated as in Eq.~\ref{eq:ce}. A $100\%$ conversion of thermal energy into kinetic energy is unlikely to happen, for example given the expected thrust losses due to the divergence of the exhaust plume. However, this is a first approach for this study, and represents the upper limit performance for this configuration. Thrust is estimated as in Eq.~\ref{eq:thrustipg}. The starting point enthalpy value of its previous characterization has been taken~\cite{tomeu}. A maximum specific plasma plume enthalpy of $h_{cal}=\SI{7.5}{\mega\joule\per{\kilo\gram}}$, at a mass flow of $\dot{m} = \SI{60}{\milli\gram\per{\second}}$, operating with air has been determined experimentally by means of the cavity calorimeter. The value of enthalpy is still low. Consequently, the dissociation degree is assumed to be respectively low, leading to the fact that the frozen losses have to be low as well.
\begin{equation}
    c_e =\sqrt{2 h_{tot}}\sim\SI{3900}{\meter\per{\second}}
    \label{eq:ce}
\end{equation}
\begin{equation}
    T=\dot{m}(h)c_e=\rho(h) v_{rel}(h) A_f \eta_c c_e\sim \SI{230}{\milli\newton}
    \label{eq:thrustipg}
\end{equation}
The precedent assumption, see Eq.~\ref{eq:ce}, leads to the following: 
\begin{itemize}
	%\item The thrust derivation currently neglects frozen losses;
	%\item The calorimeter is mounted at $\SI{2.5}{\centi\meter}$ from the IPG6-S exit;
	\item The calorimeter provides lower limit of plasma power $\rightarrow$ plasma already lost energy within IPG6-S, maximum thermal efficiency of $<25\%$ measured~\cite{dropmann};
%	\item Frozen losses are already included in the measured plasma energy;
	\item No acceleration stage is yet installed on IPG6-S, this means that only tube expansion as relevant acceleration of the plasma is present. 
\end{itemize} 
\begin{wrapfigure}{r}{.5\textwidth}
	\centering
	\vspace{-20pt}
	\includegraphics[width=6cm]{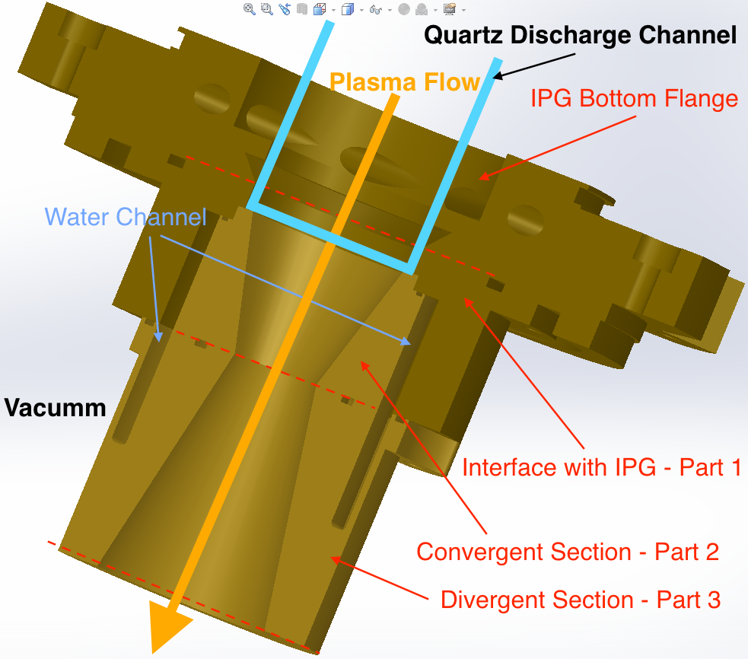}
	\caption{Cutaway of the new de Laval nozzle}
	\vspace{-40pt}
	\label{fig:nozzle}
\end{wrapfigure} Comparing this result to that of the estimated drag in Earth orbit, the possibility of full drag compensation is opened. Again, IPG6-S is not yet optimized as a thruster and there is not yet an acceleration stage applied.
Currently, a complete new facility is being completed for IPG6-S to improve reliability of thrust relevant parameters estimation. A modular de Laval water cooled nozzle has been built and it is to be tested on a new refurbished facility. It has the possibility of interchanging the convergent part for smaller/larger throat diameters and also to be used as convergent-only nozzle, this is shown in Fig.~\ref{fig:nozzle}.
It is foreseen the application of magnetic fields in the acceleration stage by means of permanent magnets and/or electromagnets. In terms of better thrust estimation, it is planned to measure exhaust velocity by a miniaturized pressure probe for radial Pitot pressure measurements.

The current outcome must be seen in terms of upper performance limit for the current IPG6-S configuration.

\section{Results}
\subsection{Thrust Estimation}
\begin{figure}[h]
    \centering
    	\subfigure[$\eta_c=0.9$]{
    	\includegraphics[width=5.7cm]{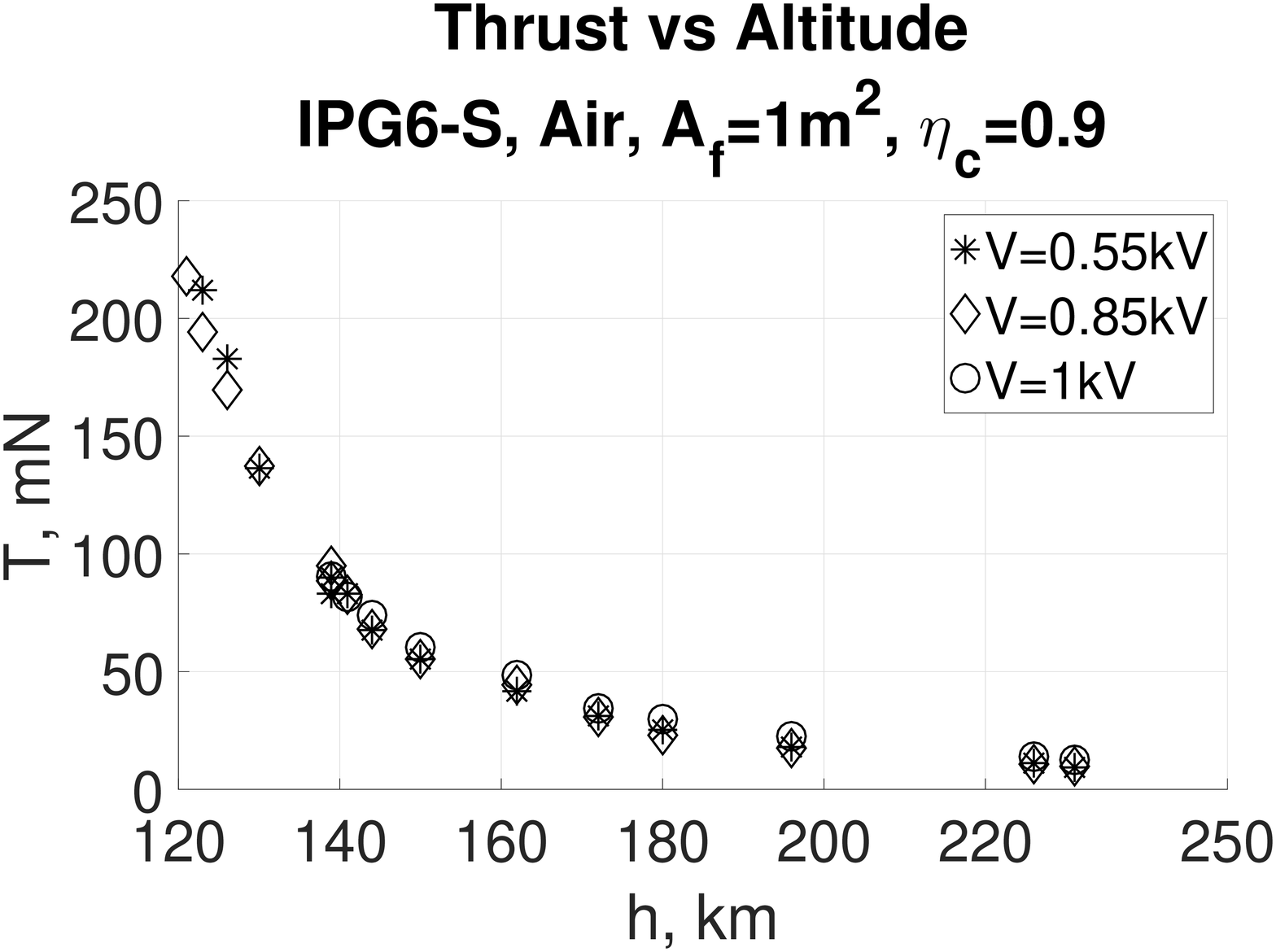}
    	\label{fig:T90A}
    	}
    	\subfigure[$\eta_c=0.46$]{
    	\includegraphics[width=5.7cm]{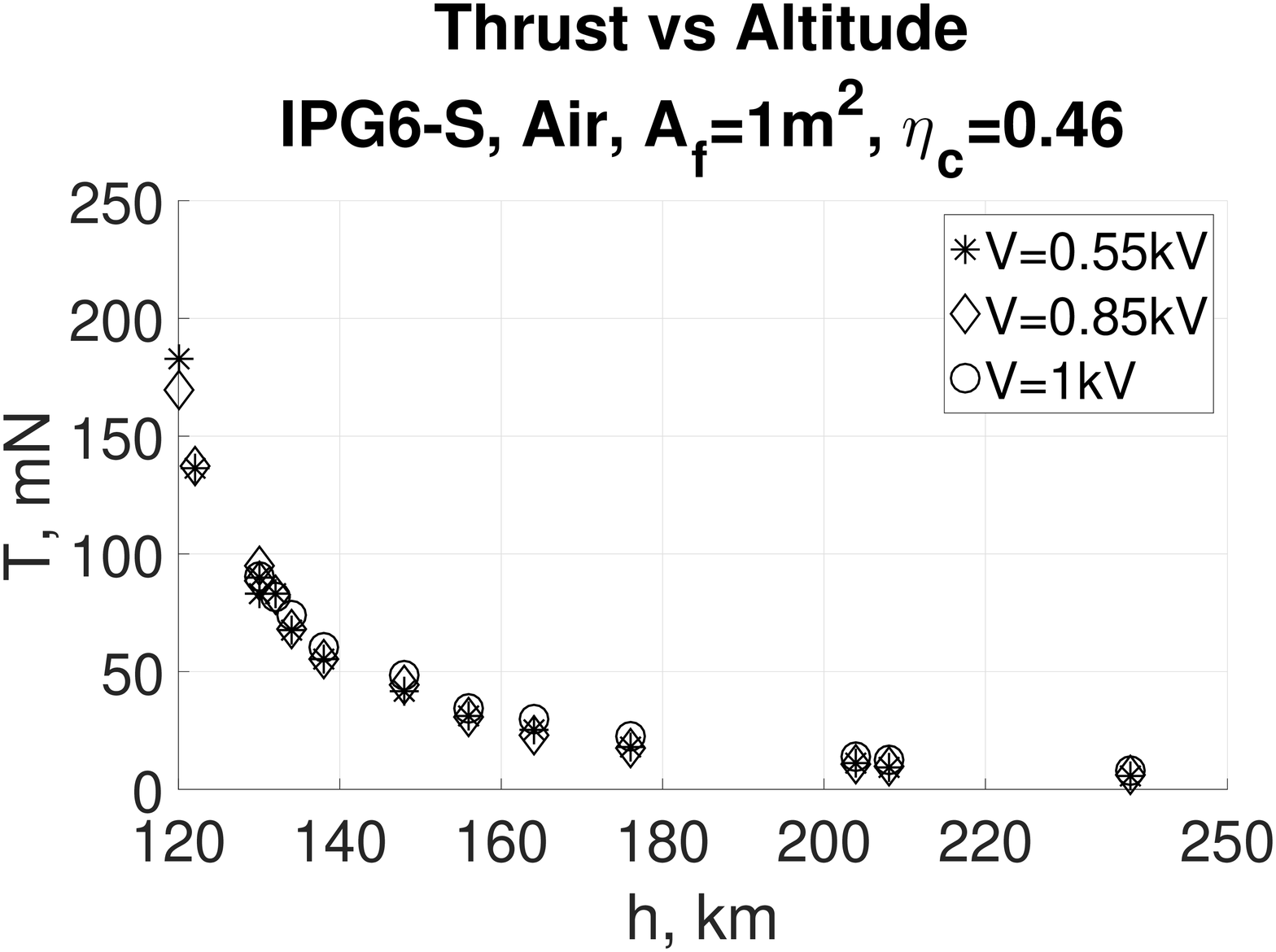}
    	\label{fig:T46A}
    	}
    	\subfigure[$\eta_c=0.35$]{
    	\includegraphics[width=5.7cm]{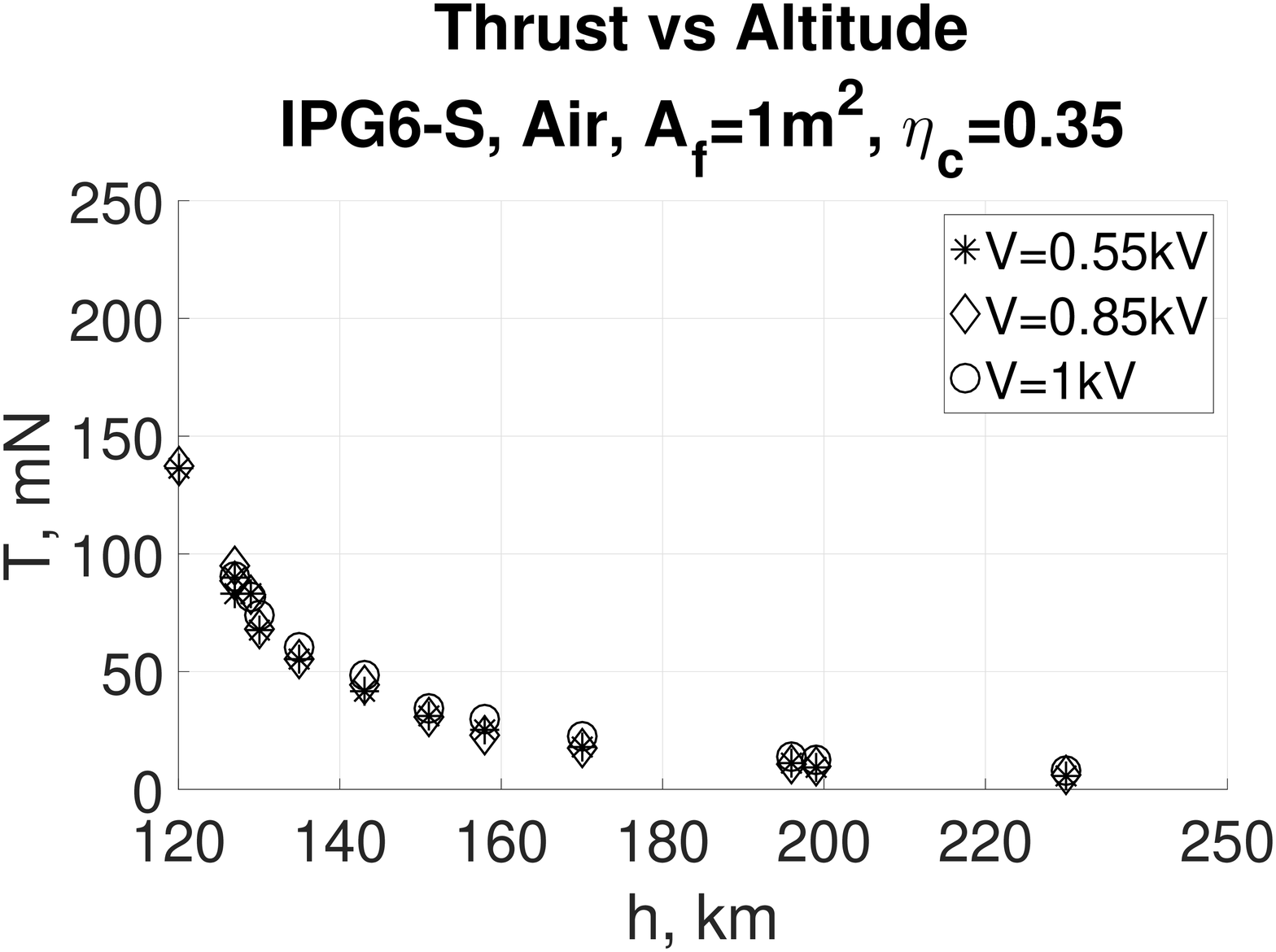}
    	\label{fig:T35A}
    }
    \caption{Upper Limit Thrust Estimation, $A_{in}=\SI{1}{\square\meter}$, Air (\ce{N2}), Avg. Solar Activity.}
    \label{fig:TA}
\end{figure}

\begin{figure}[h]
\centering
\subfigure[$\eta_c=0.9$]{
	\includegraphics[width=5.7cm]{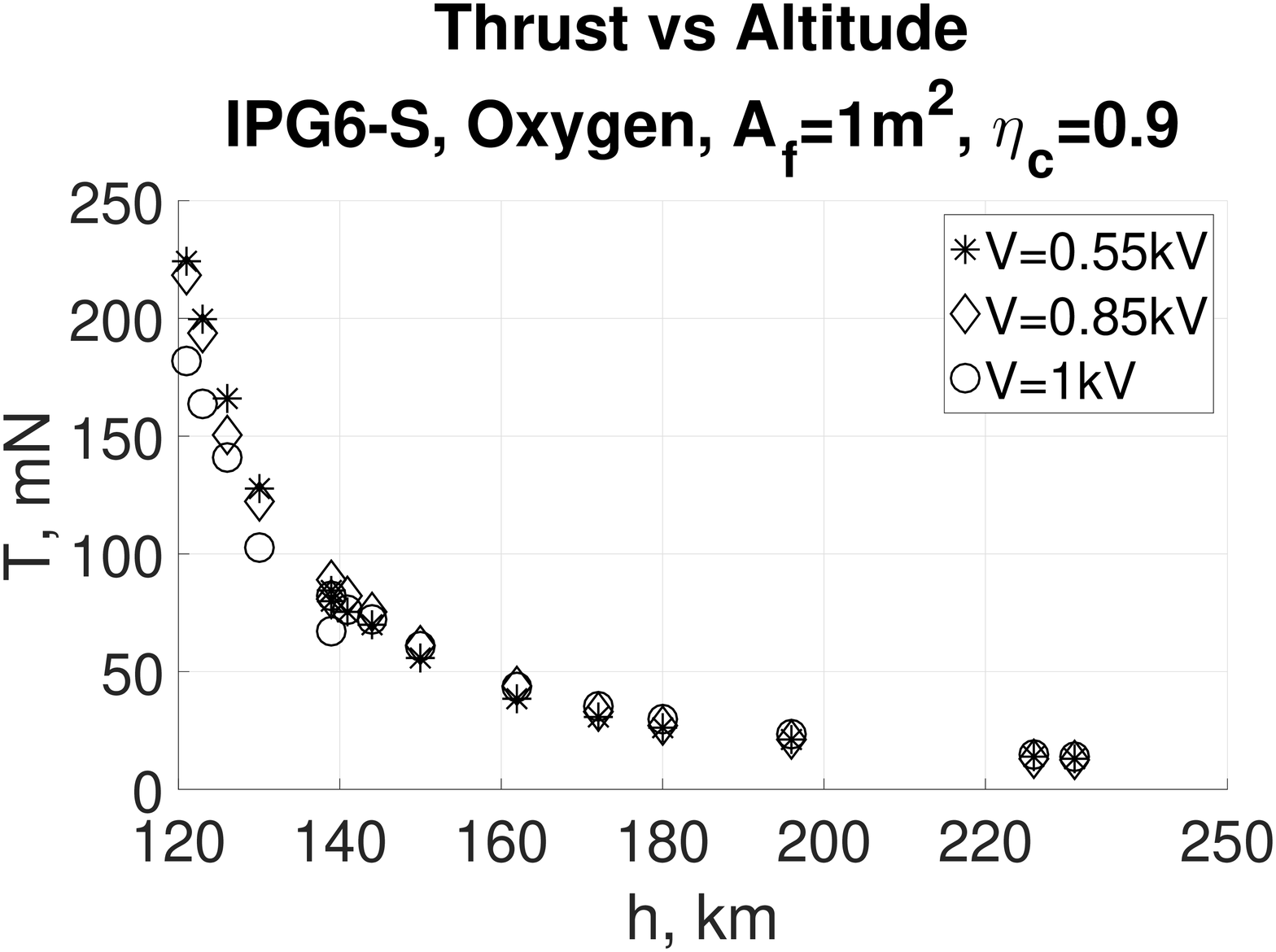}
	\label{fig:T90O}
}
\subfigure[$\eta_c=0.46$]{
	\includegraphics[width=5.7cm]{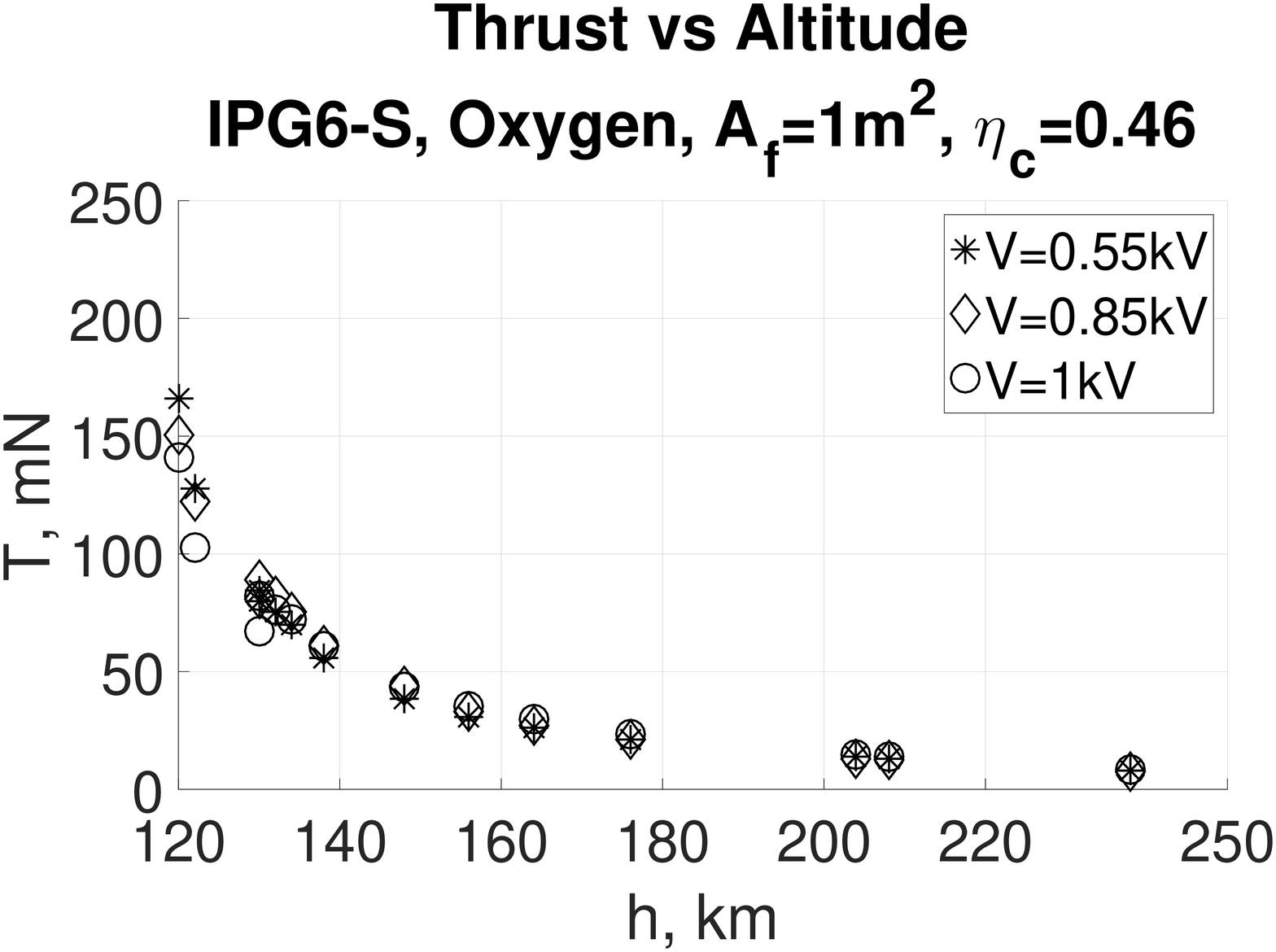}
	\label{fig:T46O}
}
\subfigure[$\eta_c=0.35$]{
	\includegraphics[width=5.7cm]{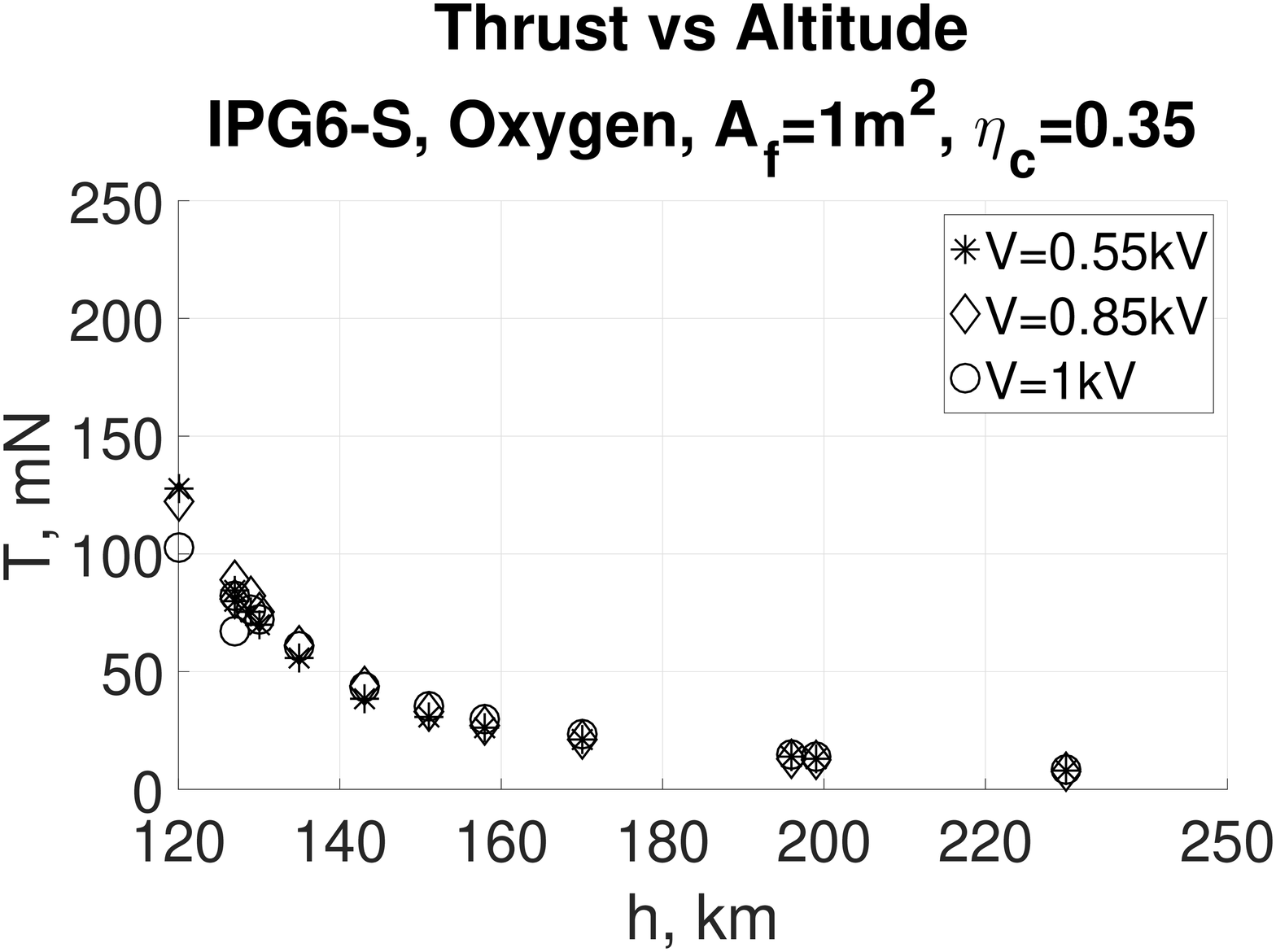}
	\label{fig:T35O}
}
\caption{Upper Limit Thrust Estimation, $A_{in}=\SI{1}{\square\meter}$, Air (\ce{O2}), Avg. Solar Activity.}
\label{fig:TO}
\end{figure}

\begin{figure}[h]
	\centering
	\subfigure[Air]{
		\includegraphics[height=7cm]{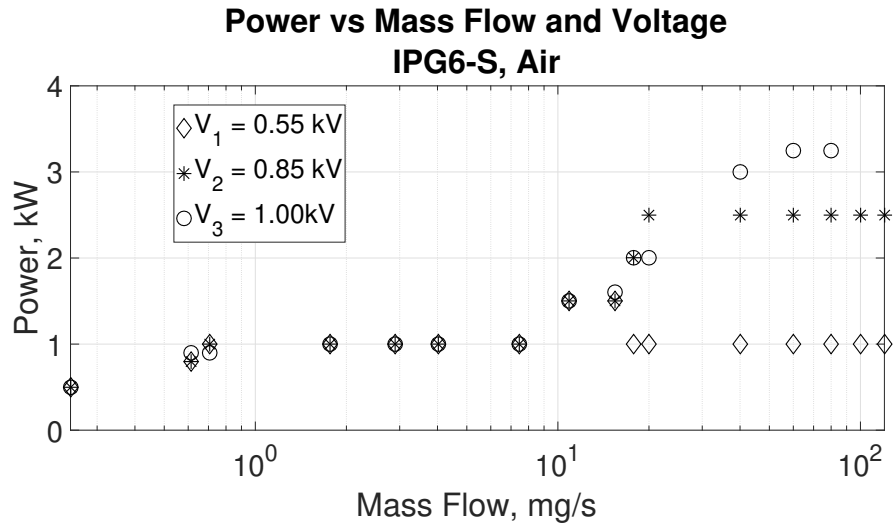}
		\label{fig:PA}
	}
	\subfigure[Oxygen]{
		\includegraphics[height=7cm]{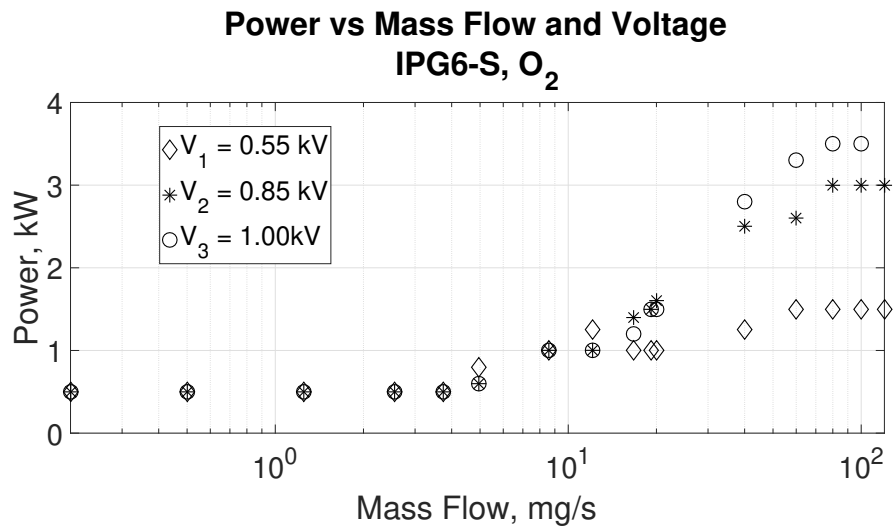}
		\label{fig:PO}
	}
	\caption{Active Power, Air and \ce{O2}}
	\label{fig:Pow}
\end{figure}

\begin{figure}[h]
	\centering
	\subfigure[Air]{
		\includegraphics[height=7cm]{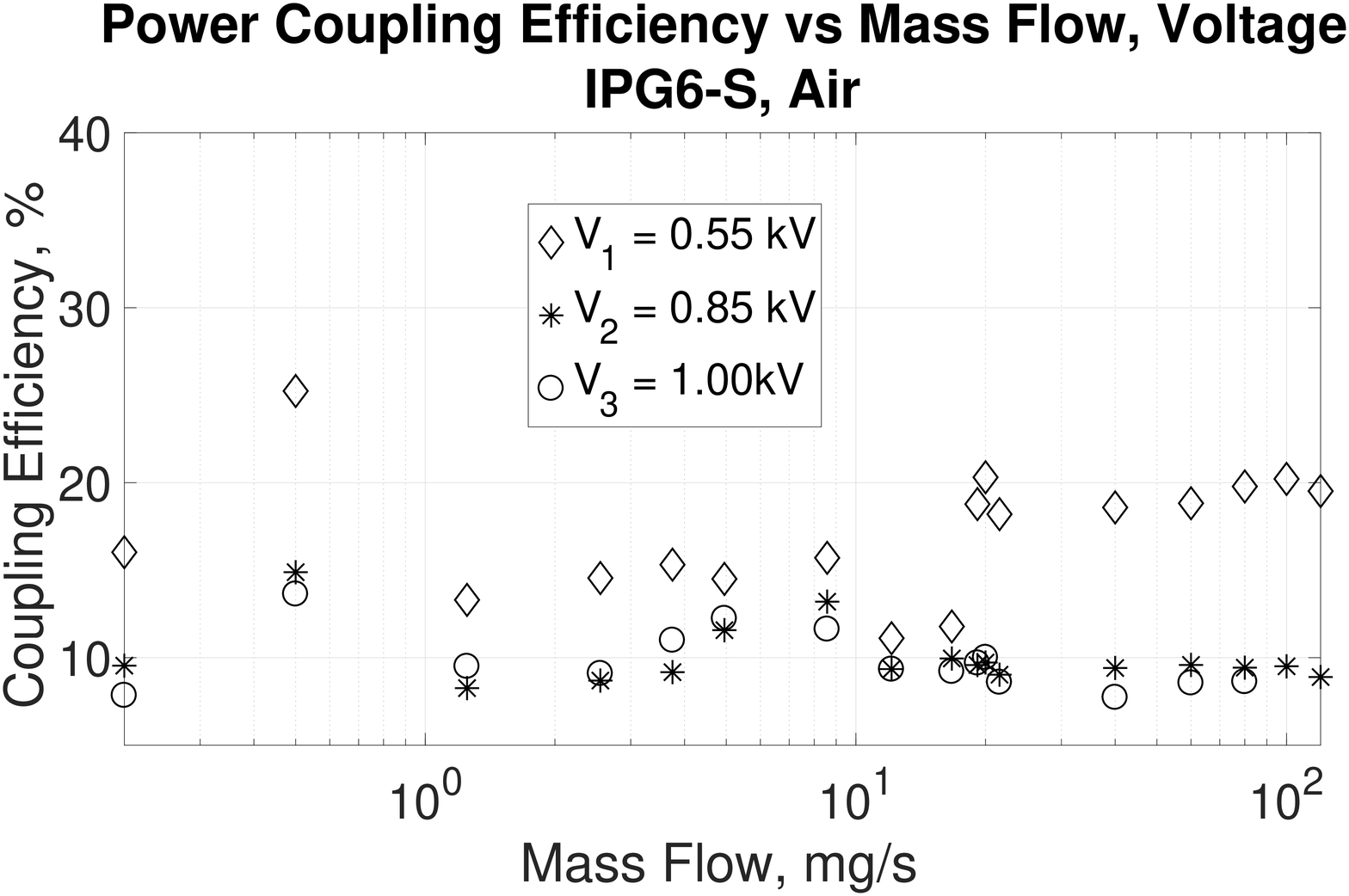}
		\label{fig:PCouplA}
	}
	\subfigure[Oxygen]{
		\includegraphics[height=7cm]{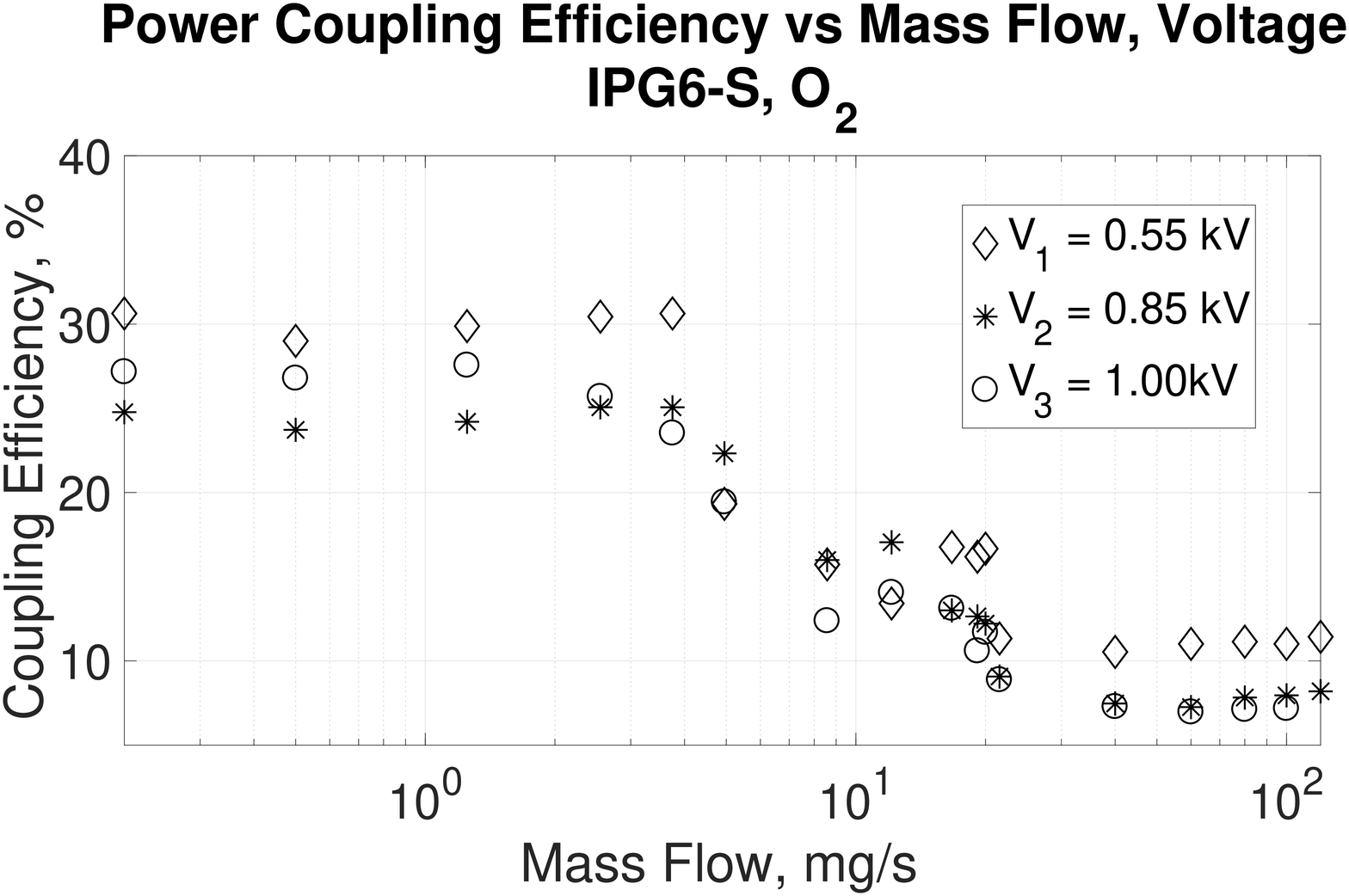}
		\label{fig:PCouplO}
	}
	\caption{Power Coupling Efficiency, Air and Oxygen}
	\label{fig:PCoupl}
\end{figure}

The estimated $\dot{m}_{thr}$ from the system analysis has been applied to IPG6-S, the enthalpy measured by the cavity calorimeter and, considering conversion of all plasma energy into kinetic energy, the calculated exhaust velocity $c_e$, see Eq.~\ref{eq:ce}, is used for thrust estimation, see Eq.~\ref{eq:thrustipg}. Thrust is plotted as a function of the $h$ in Figs.~\ref{fig:TA} and~\ref{fig:TO}, where the altitude is derived from Fig.~\ref{fig:mdot}. Thrust reaches a maximum of \SI{250}{\milli\newton} at low altitudes with \ce{O2}, slightly less for air and a minimum of \SI{5}{\milli\newton} at high altitudes for both gases. The resulting input power at the anode read from the power supply is shown in Figs.~\ref{fig:PA} and~\ref{fig:PO} for each mass flow. Missing points are due to automatic switch-off of the power supply, due to too high reflected power. Finally, in Figs.~\ref{fig:PCouplA} and~\ref{fig:PCouplO}, the power coupling efficiency, that is the ratio between calorimeter and active power for each mass flow, is shown. Fig.~\ref{fig:PCouplO} shows that the coupling efficiency is higher for lower mass flows and lower screen grid voltages, reaching the maximum of $\sim30\%$. Air shows lower efficiencies, with a maximum of $\sim25\%$ due to a higher dissociation energy. These low values can be drastically improved by means of a better thruster design. 

\newpage
\subsection{Thrust to Drag Ratio Estimation for Air and \ce{O2}}
\begin{figure}[h]
\centering
	\subfigure[$\eta_c=0.90$]{
		\includegraphics[width=9cm]{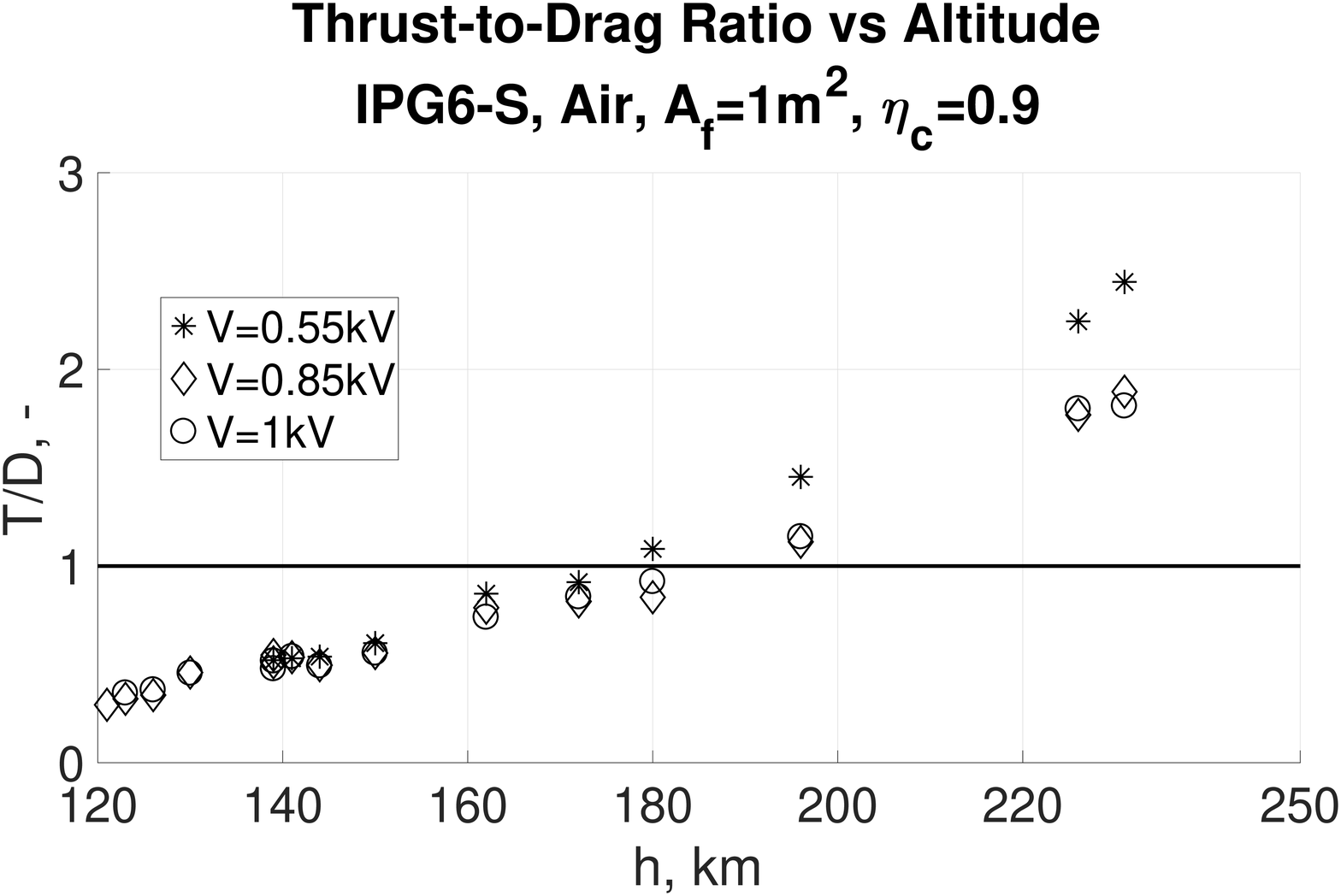}
		\label{fig:TD90A}
	}
	\subfigure[$\eta_c=0.46$]{
		\includegraphics[width=9cm]{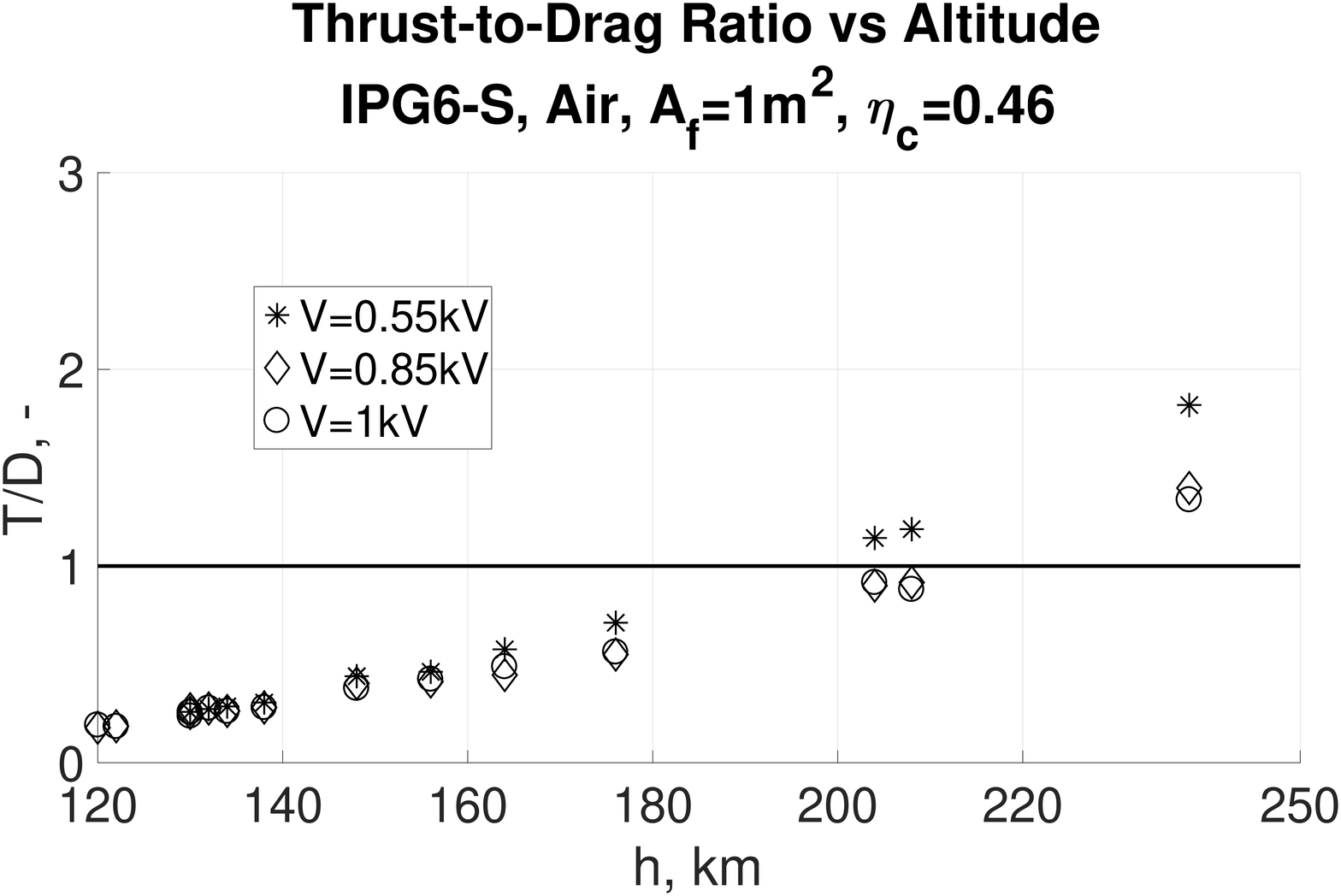}
		\label{fig:TD46A}
	}
	\subfigure[$\eta_c=0.35$]{
		\includegraphics[width=9cm]{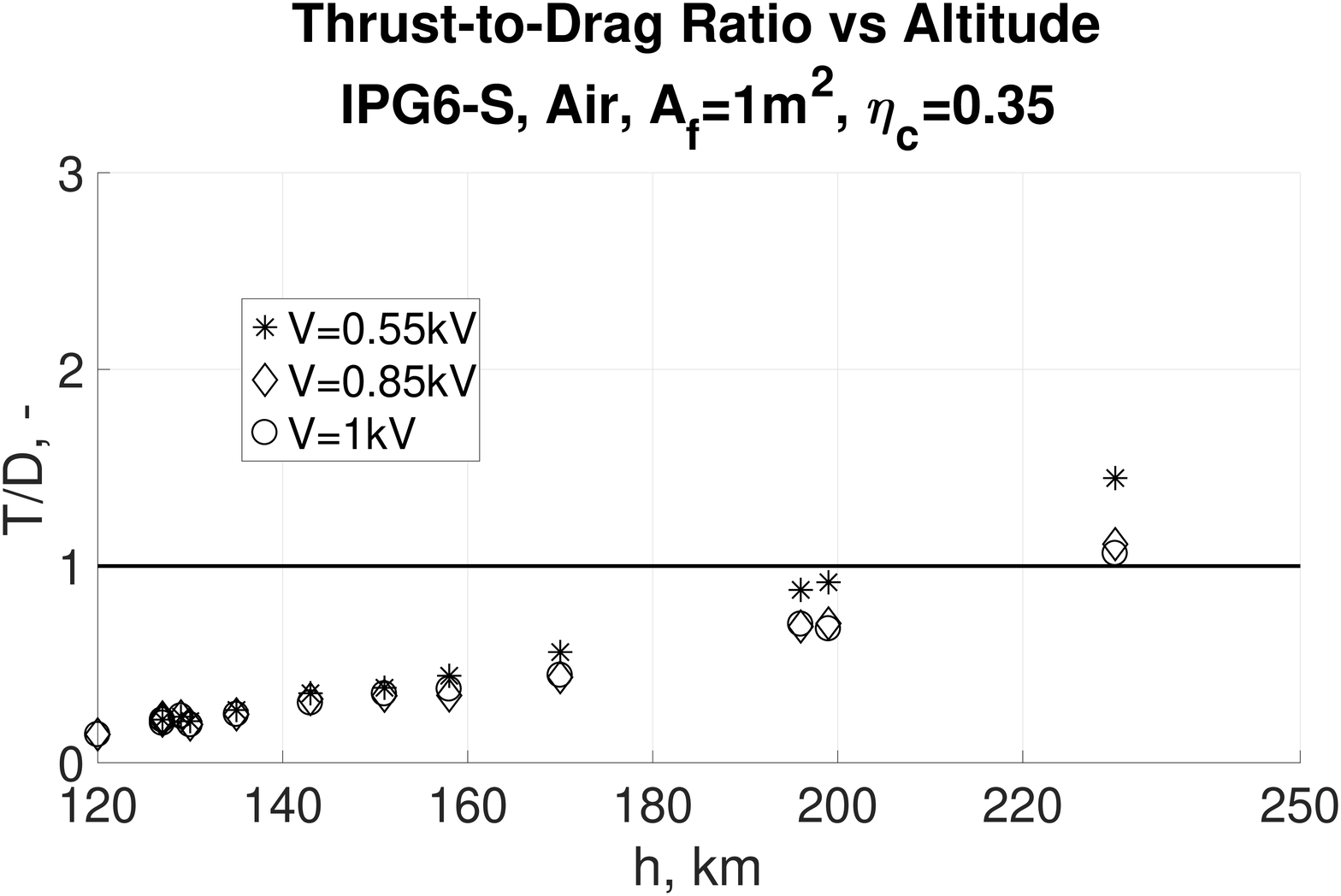}
		\label{fig:TD35A}
	}
	\caption{Thrust to Drag Ratio Estimation $A_{inlet}=\SI{1}{\square\meter}$, Air.}
	\label{fig:TDA}
\end{figure}

\begin{figure}[h]
	\centering
	\subfigure[$\eta_c=0.90$]{
		\includegraphics[width=9cm]{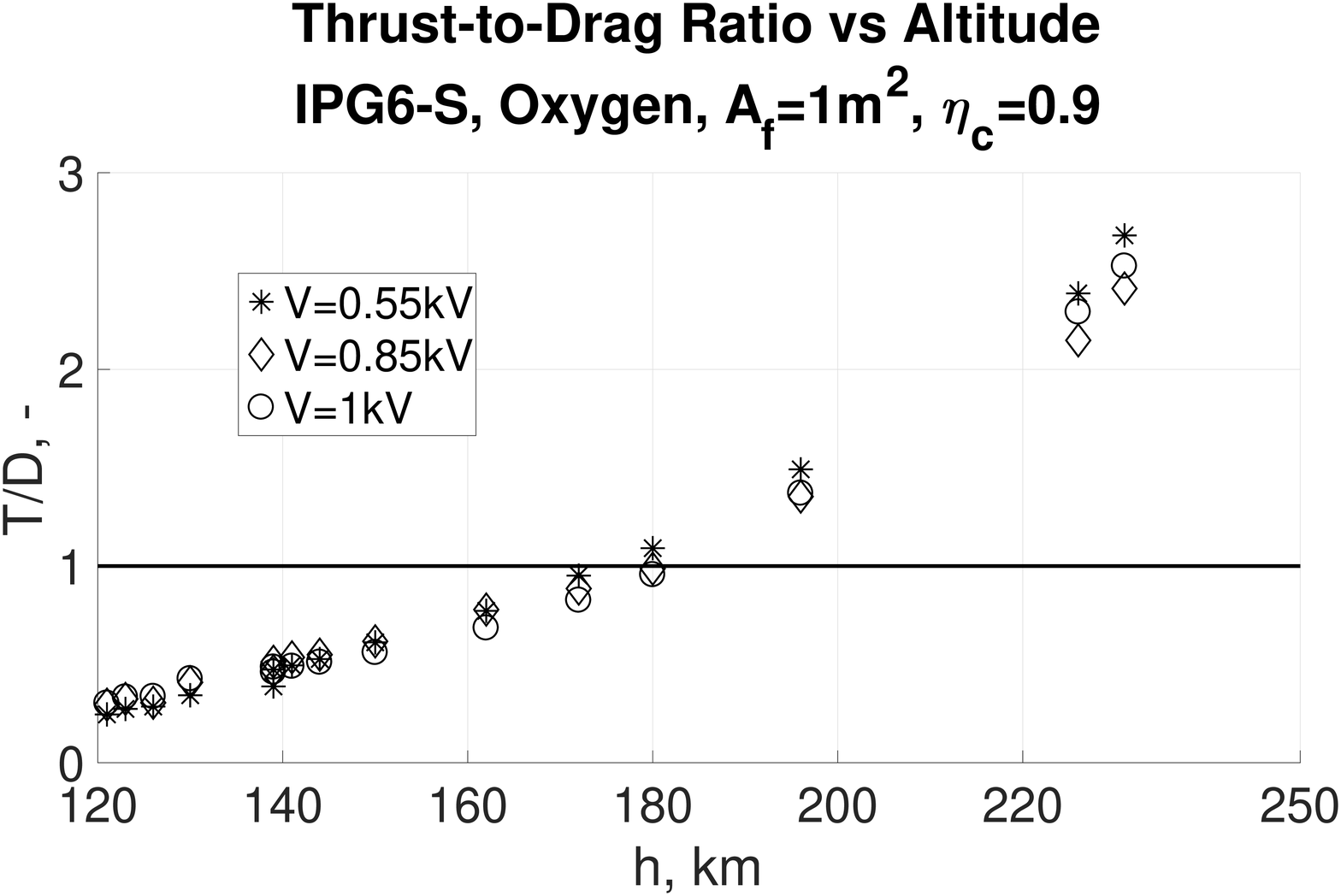}
		\label{fig:TD90O}
	}
	\subfigure[$\eta_c=0.46$]{
		\includegraphics[width=9cm]{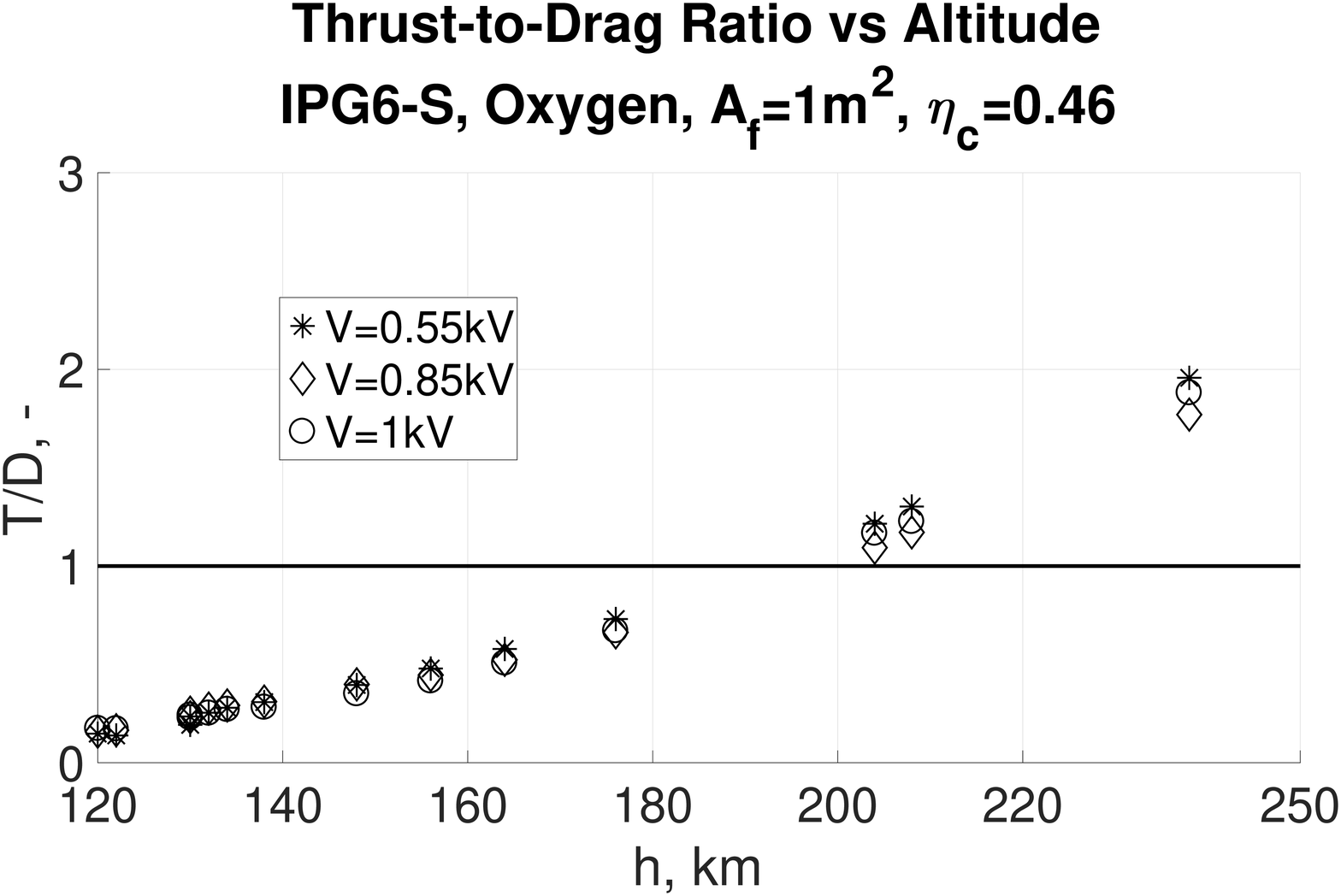}
		\label{fig:TD46O}
	}
	\subfigure[$\eta_c=0.35$]{
		\includegraphics[width=9cm]{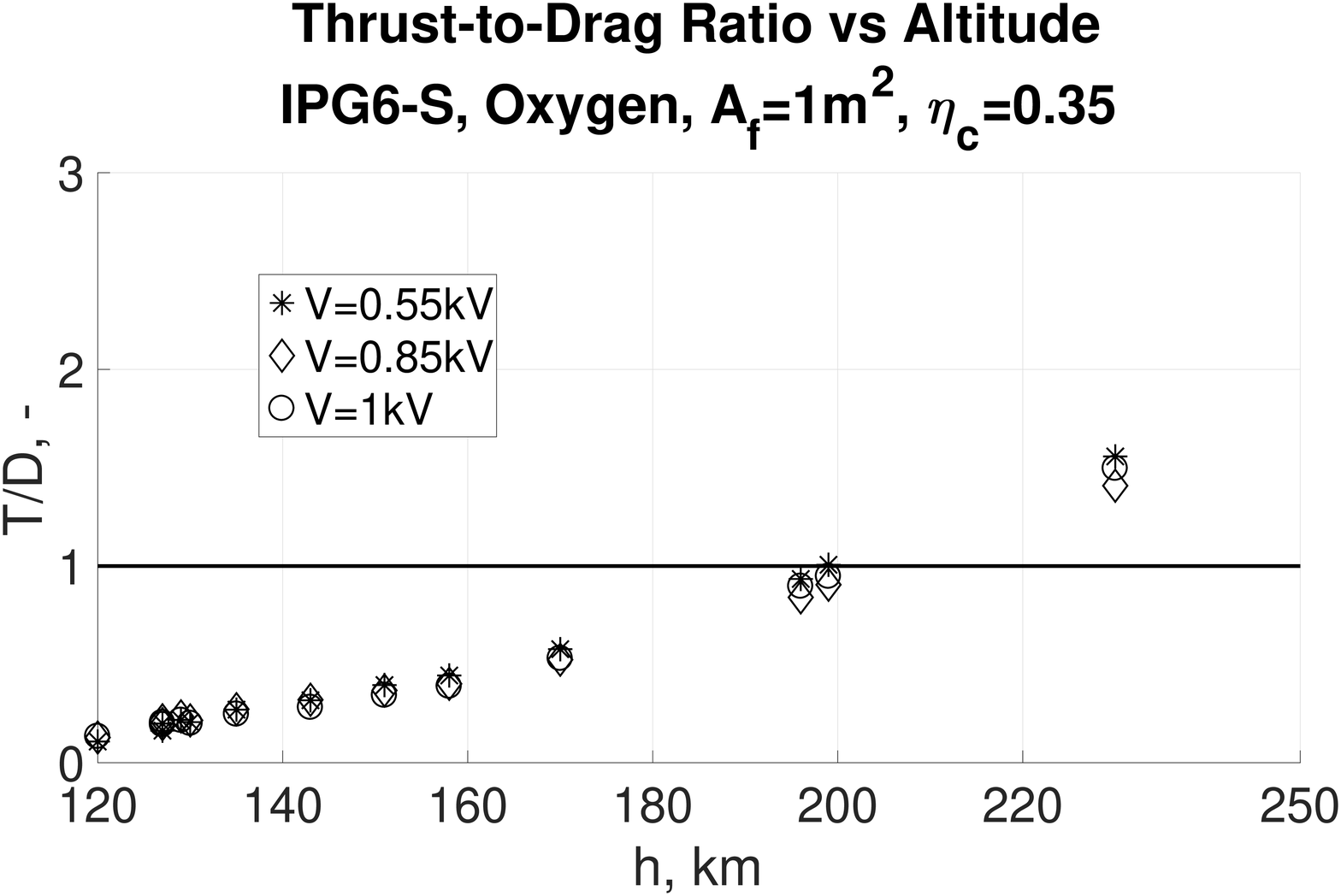}
		\label{fig:TD35O}
	}
	\caption{Thrust to Drag Ratio Estimation $A_{inlet}=\SI{1}{\square\meter}$, Oxygen.}
	\label{fig:TDO}
\end{figure}

The evaluated thrust to drag ratio is plotted as a function of the altitude for Air and \ce{O2}, for the three different screen grid voltages, $\eta_c$ and for an $A_f=\SI{1}{\square\meter}$, see Figs.~\ref{fig:TDA} and~\ref{fig:TDO}. The thrust value is divided by the drag value at the corresponding altitude in terms of mass flow. In particular it is shown that under these conditions the use of IPG6-S as thruster candidate for an ABEP system might lead to full drag compensation under the for-mentioned assumptions, over a certain altitude range.
\clearpage
%\newpage
\section{Conclusion and Outlook}
System analysis investigation set orbit and altitude ranges, collection efficiencies, input gases, drag to compensate, and power available. IPG6-S has been selected and investigated as an inductive plasma thruster candidate for an ABEP system. The facility has been improved and the generator tested with \ce{O2} and air as operating gas, for mass flows representing different altitude ranges for different intake efficiencies. Enthalpy of the plasma produced by IPG6-S has been measured through a cavity calorimeter. Subsequently, the exhaust velocity and, therefore, the thrust has been estimated. Active power reached a minimum of \SI{0.5}{\kilo\watt} and a maximum of \SI{3.5}{\kilo\watt} which is an acceptable power level for a small S/C according to the literature review. However, the power absorbed by the plasma is expected to be even lower as the cooling water absorbs most of the power~\cite{tomeu},~\cite{dropmann} this is also shown with the calculation of the power coupling efficiency shown in Fig.~\ref{fig:PCoupl}. Difference of pressure between injection and tank is not enough to achieve supersonic discharge, hence, a pump with greater suction capabilities, as well as a bigger vacuum tank, is required to obtain better simulation conditions. Three screen grid voltages have been applied for the experimental investigation, $0.55$, $0.85$ and \SI{1.00}{\kilo\volt}. Low screen grid voltages yielded higher enthalpies for low mass flows. Vice-versa high screen grid voltages yields higher enthalpies for high mass flows. Air showed better results for high mass flows, and \ce{O2} showed better results for low mass flows. At high altitudes the predominant component is \ce{O}, that means the performance of the thruster will increase by the altitude as the amount of \ce{O} will increase. The estimated thrust compared to the drag, has encouraging values. Thrust to drag ratios have been estimated and shows that such a technology might be capable of drag compensation in an ABEP application. Thrust to drag ratio has been calculated for the three selected voltages and $\eta_c$, for both \ce{O2} and Air, for an inlet area of $A_f=\SI{1}{\square\meter}$ in the ABEP altitude range. Results have shown that the thrust to drag ratio, under the for-mentioned assumptions, might be greater than one for certain altitude ranges allowing for full drag compensation. For $T/D<1$, partial drag compensation might be achieved allowing an increase of the mission lifetime.
\subsection{Outlook}
For further work, the use of a multiple stage vacuum pump and bigger tank are required for achieving better simulation conditions. Actually, a facility used for RIT testing, including a $2\times\SI{5}{\meter}$ vacuum tank is being refurbished as new experimental set-up for IPG6-S. The bigger vacuum tank and the much lower background pressure, without mass flow down to $\SI{E-4}{\pascal}$~\cite{tank7}, will allow much more reliable testing conditions. The assumption of all plasma energy converted into kinetic energy is a simplifying assumption, therefore an analysis of the acceleration strategies is required to better evaluate the exhaust velocity, hence, the thrust. As fore-mentioned, a water-cooled de Laval nozzle has been built and it will be tested to evaluate IPG6-S behaviour with an acceleration stage, providing more reliable thrust relevant parameters. The implementation, in the nozzle, of permanent magnets and/or electromagnets for plasma acceleration is currently being assessed, as the ionization level has to be evaluated. A Pitot probe will be used to estimate exhaust velocity by radial pressure measurements. A 3D S/C model is required for better estimation of S/C drag and heat loads. The realization of an ABEP S/C model for the IRS software REENT is required to evaluate S/C lifetime and orbit decay. The biggest challenge is, however, a downscaling of the IPG6-S to a suitable thruster size for ABEP application with a passive cooling system.

\section*{Acknowledgments}
F. Romano gratefully thanks the Landesgraduiertenf\"{o}rderung of the University of Stuttgart for the financial support.\\
Part of this work has been performed within the DISCOVERER project. This project has received funding from the European Union's Horizon 2020 research and innovation programme under grant agreement No 737183. This reflects only the author's view and the European Commission is not responsible for any use that may be made of the information it contains. 
\section*{References}
\bibliographystyle{elsarticle-num}
\bibliography{bibliography}

\end{document}